\newif\ifusingllncs
\newif\ifdocomments

\docommentstrue

\ifusingllncs %%% For llncs.cls
\documentclass[10pt,draft]{llncs}
\author{Norman Danner\inst{1} \and James S.~Royer\inst{2}}
\institute{%
Department of Mathematics and Computer Science, Wesleyan University,
Middletown, CT 06459, USA; Email: \email{ndanner@wesleyan.edu} \and
Department of Electrical Engineering and Computer Science, Syracuse University,
Syracuse, NY 13210, USA; Email: \email{royer@ecs.syr.edu}}

\else %%% For ndrpt.cls
\documentclass[11pt,,,]{ndrpt}
\author{Norman Danner and James S. Royer}
\address{Department of Mathematics and Computer Science, Wesleyan University,
Middletown, CT 06459, USA}
\email{ndanner@wesleyan.edu}
\address{%
Department of Electrical Engineering and Computer Science, Syracuse University,
Syracuse, NY 13210, USA}
\email{royer@ecs.syr.edu}
\thanks{\textbf{This paper is to be first published 
in \emph{Theory of Computing Systems}}, and we thank the referee for
several helpful comments.
An earlier version appears in S.B.~Cooper, B.~L\"owe, and A.~Sorbi (eds.), \emph{Computation in the Real World (Proceedings of Computability in Europe, 2007, Siena)}, vol.~4497 of \emph{Lecture Notes in Computer Science}, Springer-Verlag, Berlin, 2007.  All derivations are typeset using
\texttt{bussproofs.sty} Version~0.9, and program listings with
\texttt{listings.sty} Version~1.3.  This document is licensed under
the Creative Commons Attribution-Noncommercial~3.0 license
(\url{http://creativecommons.org/licenses/by-nc/3.0}).}
\begingroup
\catcode`$=9
\draftfoot{DRAFT{\hskip2em}\hfill \hbox{$Revision: 1.112.2.2 $, $Date: 2008-04-18 19:10:42 $}}
\endgroup
\fi

\title{Two algorithms in search of a type-system}

\ifusingllncs\usepackage{amsmath}\fi
\usepackage{amssymb}
\usepackage{bussproofs}
\EnableBpAbbreviations
\def\InfRule#1{\hbox{\textrm{#1}}}
\def\InfRuleLabel#1{\LeftLabel{\InfRule{#1}}}
\def\InfRuleP#1{\InfRule{(#1)}}
\usepackage[final]{listings}
\def\keyw#1{\ensuremath{\comb{#1}}}
\usepackage[numbers]{natbib}
\ifusingllncs
\else 
  \usepackage{ndthms}
\fi
\usepackage{suffix}

\def\eqdef{=_{\mathrm{df}}}
\let\intersect=\cap
\let\union=\cup

\let\cross=\times

\let\isom=\cong
\let\cong=\equiv

\let\equiv=\sim
\def\isomto{\mathrel{\hbox{$\to$\kern-.85em\raise1ex\hbox{{$\scriptstyle \isom$}}}}\;}
\def\ndiv{{\not \kern -.05em |\ }}

\let\comp=\circ

\def\dom{\mathrm{Dom}\;}

\def\id{\mathop{\mathrm{id}}\nolimits}

\expandafter\ifx\csname Bbb\endcsname\relax

\else

\fi

\def\setseparator{\mid}
\newcommand{\seq}[2][\relax]{
  \ifx#1\relax
    \langle#2\rangle
  \else
    \ifx#1\left
      \left\langle#2\right\rangle
    \else
      \csname #1l\endcsname\langle#2\csname #1r\endcsname\rangle
    \fi
  \fi
}

\newcommand{\set}[2][\relax]{
  \ifx#1\relax
    \{#2\}
  \else
    \ifx#1\left
      \left\{#2\right\}
    \else
      \csname #1l\endcsname\{#2\csname #1r\endcsname\}
    \fi
  \fi
}
\newcommand{\setst}[3][\relax]{\set[#1]{#2\setseparator#3}}

\def\setseparator{\mathrel{\,\stackrel{\rule{0.03em}{0.5ex}}{\rule[-.1ex]{0.03em}{0.5ex}}\,}}

\def\arrow{\mathbin{\rightarrow}}

\def\comb#1{\mathsf{#1}}

\def\eqdef{=_{\mathrm{df}}}

\def\llambda{{\lambda\hskip-.45em\lambda}}	% Semantic lambda

\def\fv{\mathop{\mathrm{fv}}\nolimits}

\def\oftype{\mathbin{:}}

\def\subst#1#2#3{#1[#2/#3]}

\def\ATR{\mathsf{ATR}}
\def\ATRm{\ATR^-}

\def\Nat{\mathsf{N}}
\def\Tally{\mathsf{T}}

\def\emptyctx{\underline{~}}
\def\typing#1#2#3#4{#1;#2\proves#3\oftype#4}
\def\ityping#1#2#3{\typing{#1}{\emptyctx}{#2}{#3}}
\def\GDtyping#1#2{\typing\Gamma\Delta{#1}{#2}}
\def\Gtyping#1#2{\ityping\Gamma{#1}{#2}}

\def\tctyping#1#2#3{#1\proves#2\oftype#3}
\def\Stctyping#1#2{\tctyping\Sigma{#1}{#2}}

\def\semanticOp#1{\mathop{\smash{\mathit{#1}}}\nolimits}
\let\combfont\mathsf
\def\comb#1{\mathop{\smash{\combfont{#1}}}\nolimits}

\def\afflambda{\lambda_r}
\def\cons{\comb{c}}
\def\destr{\comb{d}}
\def\test{\comb{t}}
\def\cond#1#2#3{\mbox{$\combfont{if}~#1~\combfont{then}~#2~\combfont{else}~#3$}}
\def\down{\comb{down}}
\def\crec{\comb{crec}}

\let\apprby\sqsubseteq
\def\apprbypot{\apprby_{\mathrm{pot}}}

\let\base\mathsf
\let\bddby\apprby
\let\bddbypot\apprbypot
\let\bmax\vee
\def\bz{\mathbf{0}}
\def\bone{\mathbf{1}}

\def\cl#1#2{#1#2}
\WithSuffix\def\cl*#1#2{(#1)#2}
\def\cost{\semanticOp{cost}}

\def\dally{\semanticOp{dally}}
\let\dmnd\Diamond

\def\Env#1{\text{$#1$-$\mathrm{Env}$}}
\def\emptyenv{[]}
\let\eps\varepsilon
\let\evalto\downarrow
\def\extend#1#2#3{#1[#2 \mapsto #3]}

\def\lh#1{|#1|}

\def\pmj{\odot}
\def\pmjb{\pmj_{\base b}}
\def\pot{\semanticOp{pot}}
\def\potden#1{\langle\!\langle#1\rangle\!\rangle}
\let\proves\vdash

\let\shiftsto\propto
\def\stdcrec{\crec\,a\,(\afflambda f.\lambda\vec v.t)}
\def\strictsubtype{\mathrel{<:}}
\let\strsubtype\strictsubtype
\def\subtype{\mathrel{\leq:}}

\def\tail{\semanticOp{tail}}
\def\tcden#1{{\|#1\|}}
\def\tccost{\semanticOp{cost}}

\def\val{\semanticOp{val}}

\ifusingllncs %%% For llncs.cls
\spnewtheorem{thm}{Theorem}{\normalfont\scshape}{\normalfont\slshape}
\spnewtheorem{lem}[thm]{Lemma}{\normalfont\scshape}{\normalfont\slshape}
\spnewtheorem{prop}[thm]{Proposition}{\normalfont\scshape}{\normalfont\slshape}
\spnewtheorem{cor}[thm]{Corollary}{\normalfont\scshape}{\normalfont\slshape}
\spnewtheorem{defn}{Definition}{\normalfont\scshape}{\normalfont}
\def\proofsketch{sketch}
\def\vec{\mathaccent"17E}
\else
\newenvironment{lem}{\begin{lemma}}{\end{lemma}}
\def\proofsketch{Proof sketch}
\fi

\ifusingllncs\else
\makeatletter
\renewcommand{\paragraph}{\@startsection%
	{paragraph}%
	{4}%
	{0in}%
	{.5\baselineskip}%
	{-.5em}%
	{\itshape}%
}
\makeatother
\fi

\let\pgmfont\mathsf

\let\contfont\mathsf
\let\tcdopfont\mathit
\let\syntaxfnfont\mathrm
\def\syntaxfn#1{\mathop{\smash{\syntaxfnfont{#1}}}\nolimits}

\def\ATS{\textit{ATS}}

\let\cat\oplus
\def\cfg#1#2#3{\langle #1,#2,\langle#3\rangle\rangle}
\WithSuffix\def\cfg*#1#2#3{\langle #1,#2,#3\rangle}

\def\cont#1{\contfont{#1}}
\def\cost{\syntaxfn{cost}}

\let\lollipop\multimap

\def\PCF{\mathsf{PCF}}
\let\phi\varphi
\def\pgm#1{\mathop{\pgmfont{#1}}\nolimits}

\def\rdp{\syntaxfn{rdp}}

\def\subst#1#2#3{#1[#3/#2]}

\def\tcdop#1{\mathop{\smash{\tcdopfont{#1}}}\nolimits}
\def\tcdcons{\tcdop{c}}
\def\tcddest{\tcdop{d}}
\def\tcdtest{\tcdop{tst}}
\def\tcdcond{\tcdop{cond}}
\def\tcddown{\tcdop{down}}
\def\tcz{\underline{\eps}}
\let\transto\rightsquigarrow
\let\trevalto\downharpoonleft

\def\unknown{\cl ?\emptyenv}

\ifusingllncs\else\usepackage[final]{hyperref}\fi

\begin{document}
\ifusingllncs
	\pagestyle{myheadings}\markright{\thepage}
\fi

\maketitle

\begin{abstract} % 200 words max --- also no references used
  The authors' $\ATR$ programming formalism is a version of
  call-by-value $\PCF$ under a complexity-theoretically motivated type
  system.  
  $\ATR$ programs run in type-$2$ polynomial-time and all
  standard type-$2$ basic feasible functionals are
  $\ATR$-definable
  ($\ATR$ types are confined to levels  $0$, $1$, and~$2$).
  A limitation of the original version of $\ATR$ is that
  the only directly expressible recursions are
  tail-recursions.  Here we extend $\ATR$ so that a broad range
  of affine recursions are directly expressible.  In particular,
  the revised $\ATR$ can fairly naturally express
  the classic insertion- and selection-sort algorithms,
  thus overcoming a sticking point of most prior
  implicit-complexity-based formalisms.
  The paper's main work is in refining the
  original time-complexity semantics for $\ATR$ to show that 
  these new recursion schemes do not lead out of the realm of
  feasibility.
\end{abstract}

\section{Introduction}

\subsection{Feasible programming and Affine Tiered Recursion}

As \citet{Hofmann:non-size-incr} has noted, a problem with 
implicit characterizations of complexity classes is that they
often fail to capture many natural \emph{algorithms}---usually
because the complexity-theoretic types used to control primitive
recursion impose draconian
restrictions on programming.  For example, in Bellantoni and
Cook's \citep{Bellantoni-Cook:Recursion-theoretic-char} and Leivant's
\citep{Leivant:Ram-rec-I} well-known characterizations of the
poly\-nom\-ial-time computable functions, 
a value that is the result of a recursive call cannot itself be
used to drive a recursion.
But, for instance, the
recursion clause of insertion-sort has the form $\pgm{ins\_sort}(\pgm{cons}(a,
l)) = \pgm{insert}(a, \pgm{ins\_sort}(l))$, where $\pgm{insert}$ is defined by
recursion on its second argument; selection-sort presents analogous
problems.

\citet{Hofmann:non-size-incr,Hofmann:ic03} addresses this problem 
by noting that the output of a non-size-increasing program 
(such as $\pgm{ins\_sort}$) 
can be safely used to drive another recursion, as it
cannot cause the sort of complexity blow-up the B-C-L restrictions
guard against.
To incorporate such
recursions, Hofmann defines a higher-order language with typical first-order
types and a special type~$\Diamond$ through which
functions defined recursively 
must ``pay'' for any use of size-increasing constructors, in effect
guaranteeing that there is no size increase.  Through this scheme Hofmann is 
able
to implement many natural algorithms while still ensuring that any typable
program is non-size-increasing
poly\-nom\-ial-time computable (\citet{aehlig-schw:tocl02} 
sketch an extension that
captures all of polynomial-time).  

Our earlier paper~\citep{danner-royer:ats,danner-royer:ats-lmcs}, 
hereafter referred to as~\ATS,
takes a different approach to constructing a usable programming
language with guaranteed resource usage.  \ATS\ introduces a type-$2$
programming formalism called $\ATR$, for \emph{Affine Tiered Recursion},
based on call-by-value $\PCF$ for which the underlying model of computation
(and complexity) is a standard abstract machine.\footnote{In our 
earlier~\citep{danner-royer:ats} 
$\ATR$ stood for Affine Tail Recursion; we re-christened it 
in~\citep{danner-royer:ats-lmcs}.}
$\ATR$'s type system comes in two parts:
one that is motivated by the tiering and safe/normal notions of
\cite{Leivant:Ram-rec-I} and
\cite{Bellantoni-Cook:Recursion-theoretic-char} and serves to control the
size of objects, and one that is motivated by notions of affine-ness that
serves to control time.  Instead of restricting to primitive recursion, $\ATR$
has an operator for recursive definitions; affine types and explicit clocking
on the operator work together to prevent any complexity blow-up.
In \ATS\ we give a denotational semantics to $\ATR$ types and terms in which the
size restrictions play a key part.  This allows us, for example, to
give an~$\ATR$ \emph{definition} 
of a primitive-recursion-on-notation combinator
(\emph{without} explicit bounding terms) that
preserves feasibility.
We also give a \emph{time-complexity semantics} and
use it to prove that each type-$2$ $\ATR$ program has a (second-order)
polynomial run-time.%
\footnote{
These kinds of results may also have applications in the type of
static analysis for time-complexity that
\citet{frederiksen-jones:recognition} investigate.}
Finally, we show that the type-$2$ basic feasible functionals
(an extension of polynomial-time computability to type-$2$) 
of \citet{mehlhorn:stoc74} and \citet{cook-urquhart:fca} are
$\ATR$ definable.
However, the version of~$\ATR$ defined in~\ATS\
is still somewhat limited as its only base type is binary
words and the only recursions allowed are tail-recursions.

\subsection{What is new in this paper}
In this paper we extend $\ATR$ to encompass a broad class of feasible affine
recursions.  We demonstrate these extensions by giving fairly
direct and natural versions of insertion- and selection-sorts on lists
(Section~\ref{sec:programming})\footnote{We discuss quick-sort
in Section~\ref{sec:concl}.}
as well as the primitive-recursion-on-notation combinator
(in Section~\ref{sec:polynomial-bounds}).
As additional evidence of $\ATR$'s support for programming we do not
add lists as a base type, but instead show how to implement them
over $\ATR$'s base type of binary words.

The ``two algorithms'' of the title should not be interpreted as
referring to insertion- and selection-sort, but rather the 
recursion schemes that those two algorithms exemplify.  Most implicit 
characterizations restrict to structural recursion, resulting in
somewhat ad-hoc implementations of other kinds of recursion by simulation.
We chose insertion-
and selection-sort for our prime examples in this paper because
they embody key forms non-structural one-use recursion; we capture these
key forms in what we call \emph{plain affine recursion}.
We feel that by handling any plain affine recursive program,
we have shown that our system
can deal with almost all standard feasible linear recursions.

The technical core of this paper is the extension of the Soundness Theorem
from~\ATS\ (which handled only tail recursions) to the current version
of~$\ATR$.  After defining an evaluation semantics in
Section~\ref{sec:atr-formalism} 
and surveying and simplifying the \emph{time-complexity
semantics} of~\ATS\ in Section~\ref{sec:t-c-survey}, we introduce
and prove the Soundess Theorem for \emph{plain affine recursions} in
Section~\ref{sec:atr-soundness}.  In
Section~\ref{sec:polynomial-bounds} we use the Soundness Theorem to
relate $\ATR$-computable functions to the type-$2$ basic feasible functions.
Since plain affine recursions include those
used to implement lists and the sorting algorithms, this significantly
extends our original formalism to the point where many standard algorithms
can be naturally expressed while ensuring that we do not leave the
realm of type-$2$ feasibility (and in particular, polynomial-time for
type~$1$ programs).

With the exception of the \InfRuleP{Shift} typing rule, we provide
full definitions of all terms in this paper, and we believe that it
can be understood on its own.  However, the paper is not entirely
self-contained: some of the proofs are
adaptations of corresponding proofs in \ATS, and in those cases
we refer the reader to that paper for details.

\subsection{Acknowledgment}
Part of the motivation for this paper was a challenge
to give natural versions of insertion-, selection-, and quick-sorts
within an implicit complexity formalism issued by Harry Mairson in
a conversation with the second-author.

\section{The~$\ATR$ formalism}
\label{sec:atr-formalism}

\subsection{Types, expressions, and typing}
\label{sec:atr-expressions-etc}
An $\ATR$ base type has the form $\Nat_L$, where \emph{labels}
$L$ are elements of the set
$(\Box\dmnd)^*\bigcup \dmnd(\Box\dmnd)^*$ (our use of~$\dmnd$
is unrelated to Hofmann's); the intended interpretation
of $\Nat_L$ is $K\eqdef\set{\bz,\bone}^*$.
The labels are ordered by
$\eps\leq\dmnd\leq\Box\dmnd\leq\dmnd\Box\dmnd\leq\dotsb\,$.
We define a subtype relation on the base types by
$\Nat_L\subtype\Nat_{L'}$ if $L\leq L'$
and extend it to function
types in the standard way.
Roughly, we can think of type-$\Nat_\varepsilon$ values as 
basic string inputs, type-$\Nat_\dmnd$ values as the result
of poly\-nom\-ial-time computations over $\Nat_\varepsilon$-values,
type-$\Nat_{\Box\dmnd}$-values as the result applying an oracle
(a type-1 input) to $\Nat_\dmnd$-values, type-$\Nat_{\dmnd\Box\dmnd}$ 
values as the result of poly\-nom\-ial-time computations over 
$\Nat_{\Box\dmnd}$-values, etc.
To make an analogy with the safe/normal distinction of
\citet{Bellantoni-Cook:Recursion-theoretic-char}, oracular types
correspond to normal arguments 
and computational types correspond
to safe arguments
(once we apply an oracle, we ``reset'' our
notion of what constitutes potentially large data---but we do not ``flatten''
the notion by having one oracular and one computational type).
$\ATR$'s denotational semantics works to enforce these intuitions.
$\Nat_L$ is called an \emph{oracular} (respectively,
\emph{computational}) type when $L\in (\Box\dmnd)^*$ 
(respectively, $\dmnd(\Box\dmnd)^*$). 
We let $\base b$ (possibly decorated) range over base types.
Function types are formed as usual from the base types.
We sometimes write $(\sigma_1,\dots,\sigma_k)\arrow\sigma$ or
$\vec\sigma\arrow\sigma$ for $\sigma_1\arrow\dots\arrow\sigma_k\arrow\sigma$.

\begin{defn}
For any type~$\sigma$ define $\tail(\sigma)$ by
$\tail(\base b) = \base b$ and $\tail(\sigma\arrow\tau) = \tail(\tau)$.
\end{defn}

\begin{defn}
\label{defn:predicative-etc}
A type~$\sigma$ is \emph{predicative} when $\sigma$ is a base type
or when $\sigma = \sigma_1\arrow\dots\arrow\sigma_k\arrow\Nat_L$
and $\tail(\sigma_i)\subtype \Nat_L$ for all~$i$.  A type
is \emph{impredicative} if it is not predicative.  A
(function) type $\sigma_1\arrow\dots\arrow\sigma_k\arrow\Nat_L$ is
\emph{flat} if $\tail(\sigma_i) = \Nat_L$ for some~$i$.
A type is \emph{strict} if it is not flat.
\end{defn}

The interpretation of the arrow types entails a significant amount
of work in the semantics, which we do in~\ATS.  Very briefly, our
semantics takes seriously the size information implicit in the
labeled base types.  In particular, the full type structure
is ``pruned'' to create what we call the \emph{well-tempered semantics}
so that the function spaces of flat and impredicative types
consist only of functions with appropriate growth rates.
The relevant points are the following:
\begin{enumerate}
\item If $f\oftype(\sigma_1,\dots,\sigma_k)\arrow\base b$ and
	$\base b \subtype \tail(\sigma_i)$, then $\lh f$ is bounded by a 
	safe polynomial (see Definition~\ref{defn:safety}), where
	$\lh f$ measures the growth rate of~$f$ and is defined
	in Definition~\ref{defn:function-length}.
\item As a special case of the previous point,
	if $f\oftype(\sigma_1,\dots,\sigma_k)\arrow\base b$ and
	$\base b\strsubtype\tail(\sigma_i)$, then $\lh f$ is independent
	of its $i$-th argument.
\item Recursive definitions in~$\ATR$ typically have flat types; the
	restriction on growth rates ensures that such recursively-defined
	functions do not lead us out of the realm of feasibility.
\end{enumerate}
As this paper
is concerned primarily with syntactic matters (extending the allowable
forms of recursions), we do not go into full details of the 
denotational semantics here, instead referring the reader to 
Sections~6--9 of~\ATS.

The $\ATR$ expressions are defined in Figure~\ref{fig:expr}.
We use $v, x, y, z$
for variables, $a$ for elements of~$K$, $\alpha$, $\beta$ for oracles,
and $t$ for expressions
(all possibly sub- and super-scripted and with primes). 
We can think of oracle symbols as external function calls.
Formally, they are constant symbols for elements of the 
$\ATR$-type structure with type-level~$1$; 
as such, each oracle symbol is assumed to be labeled with its type,
which we write as a superscript when it needs to be indicated.%
\footnote{As a constant, an oracle symbol is closed, and we will suppress
the interpretation of oracle symbols in the semantics.}
The more-or-less typical expression-forming operations correspond
to adding and deleting a left-most bit ($\cons_0$, $\cons_1$, and $\destr$),
testing whether a word begins with a $\bz$ or a $\bone$ ($\test_0$ and $\test_1$),
and a conditional.
The intended interpretation of $\down s\,t$ is a length test that
evaluates to
$s$ when $\lh s\leq\lh t$ and $\eps$ when $\lh{s}>\lh{t}$.  The recursion
operator is $\crec$, standing for \emph{clocked recursion}.
In Section~\ref{sec:programming} we present several sample~$\ATR$
programs.%
%%%
%%% ATR expressions.
%%%
\begin{figure}[tb]
\begin{align*}
K &::= \set{\bz,\bone}^* \\
E &::= V \mid O \mid K \mid \lambda V.E \mid EE \\
  &\qquad\mid \cons_0 E \mid \cons_1 E \mid \destr E \mid \test_0 E \mid \test_1 E\mid \cond{E}{E}{E} \\
  &\qquad\mid\down E\,E \mid \crec K(\afflambda V.E)
\end{align*}
\caption{$\ATR$ expressions.  $V$ is a set of variable symbols and
$O$ a set of oracle symbols.\label{fig:expr}}
\end{figure}

The typing rules are given in Figure~\ref{fig:typing}.
Type contexts are split into intuitionistic and affine zones as with
Barber and Plotkin's DILL~\citep{barber:dill}.
When we write $\Delta_0\union\Delta_1$ we implicitly assume that
the environments are consistent (i.e., assign the same type to variables
in $\dom\Delta_0\intersect\dom\Delta_1$) and when we write
$\Delta_0,\Delta_1$ we implicitly assume that the environments have
disjoint domains.
Variables in the intuitionistic zone correspond 
to the usual $\arrow$ introduction and elimination rules and variables
in the affine zone are intended to be recursively defined; variables that occur
in the affine zone are said to \emph{occur affinely} in the judgment. 
The \InfRule{$\crec$-I} rule 
serves as both introduction and elimination
rule for the implicit $\lollipop$~types (in the rule 
$\vec{\base b} = \base b_1,\dots,\base b_k$ and
$\vec v\oftype\vec{\base b}$ stands for $v_1\oftype \base b_1,\dots,v_k\oftype \base b_k$).
We use $\afflambda$ as the abstraction operator for variables introduced
from the affine zone of the type context to further distinguish them from
intuitionistic variables.
The typing rules enforce a ``one-use'' restriction on affine variables
that we discuss in Section~\ref{sec:plain-affine-recursion}.
Forbidding affine variables in the conditional test is primarily
a convenience and can be easily worked around with \keyw{let}-bindings.
Two of the inference rules come with side-conditions:
\begin{description}
\item[\InfRuleP{$\crec$-I} side-condition]
If $\base b_i\subtype\base b_1$ 
then $\base b_i$ is oracular (including $i=0$).  
\item[\InfRuleP{$\arrow$-E} side-condition]
At most one of $\Delta_0$ and $\Delta_1$ is non-empty, and if
$\Delta_1$ is non-empty then $\sigma$ is a base type.
\end{description}
Recalling our analogy
of oracular types with normal arguments, the \InfRuleP{$\crec$-I}
side-condition says that the
clock bound (the first argument in a recursive definition) is normal
and its size only depends on normal data.  Thus, while the clock bound
can be changed during a recursive step, this change is well-controlled.
This is the core of the Termination Lemma (Theorem~\ref{termination-lemma}),
in which we
prove a polynomial size-bound on the growth of the arguments to~$f$,
which in turn allows us to prove such bounds on all terms.
The intuition behind the \InfRuleP{$\arrow$-E} side-condition
is that an affine variable~$f$ 
may occur in either the operator or argument
of an application, but not both.  Furthermore, if it occurs in the argument,
then it must be a ``completed'' application in order to prevent the
operator from duplicating it (our call-by-value semantics will thus
recursively evaluate this complete application once and then plug the
result into the operator).
%%%
%%% ATR typing.
%%%
\begin{figure}[tb]
\begin{tabular}{lr}
\multicolumn{2}{c}{%
$%
\AXC{}
\InfRuleLabel{Zero-I}
\UIC{$\GDtyping\eps{\Nat_\eps}$}
\DisplayProof
$%
\quad
$%
\AXC{}
\InfRuleLabel{Const-I}
\UIC{$\GDtyping a {\Nat_{\dmnd}}$}
\DisplayProof
$%
\quad
$
\AXC{}
\InfRuleLabel{Oracle-I}
\UIC{$\GDtyping {\alpha^\sigma}\sigma$}
\DisplayProof
$
}
\\[1.5ex]
$%
\AXC{}
\InfRuleLabel{Int-Id-I}
\UIC{$\typing{\Gamma,v\oftype\sigma}\Delta v\sigma$}
\DisplayProof
$%
&
$%
\AXC{}
\InfRuleLabel{Aff-Id-I}
\UIC{$\typing\Gamma{\Delta,v\oftype\sigma}v\sigma$}
\DisplayProof
$
\\[2ex]
$%
\AXC{$\ityping\Gamma s\sigma$}
\InfRuleLabel{Shift}
\RightLabel{($\sigma\shiftsto\tau$)}
\UIC{$\ityping\Gamma s\tau$}
\DisplayProof
$%
&
$%
\AXC{$\GDtyping s\sigma$}
\InfRuleLabel{Subsumption}
\RightLabel{($\sigma\subtype\tau$)}
\UIC{$\GDtyping s\tau$}
\DisplayProof
$
\\[3ex]
\multicolumn{2}{c}{%
$%
\AXC{$\GDtyping s{\Nat_{\dmnd_d}}$}
\InfRuleLabel{$\cons_a$-I}
\UIC{$\GDtyping {(\cons_a s)} {\Nat_{\dmnd_d}}$}
\DisplayProof
\qquad
\AXC{$\GDtyping s {\Nat_L}$}
\InfRuleLabel{$\destr$-I}
\UIC{$\GDtyping {\destr s} {\Nat_L}$}
\DisplayProof
\qquad
\AXC{$\GDtyping s {\Nat_L}$}
\InfRuleLabel{$\test_a$-I}
\UIC{$\GDtyping {\test_a s} {\Nat_L}$}
\DisplayProof
$%
}
\\[2ex]
\multicolumn{2}{c}{%
\AXC{$\typing\Gamma{\Delta} s{\Nat_{L_0}}$}
\AXC{$\ityping\Gamma t{\Nat_{L_1}}$}
\InfRuleLabel{$\down$-I}
\BIC{$\typing\Gamma{\Delta}{(\down st)}{\Nat_{L_1}}$}
\DisplayProof
}
\\[2.5ex]
\multicolumn{2}{c}{%
\AXC{$\Gtyping s{\Nat_{L}}$}
\AXC{$\typing\Gamma{\Delta_0} {t_0}{\Nat_{L'}}$}
\AXC{$\typing\Gamma{\Delta_1} {t_1}{\Nat_{L'}}$}
\InfRuleLabel{$\comb{if}$-I}
\TIC{$\typing\Gamma{\Delta_0\union\Delta_1}{(\cond s{t_0}{t_1})}{\Nat_{L'}}$}
\DisplayProof
}
\\[3ex]
\multicolumn{2}{c}{%
\AXC{$\typing{\underline{~}}{\underline{~}}a{\Nat_\dmnd}$}
\AXC{$\typing{\Gamma,\vec v\oftype\vec{\base b}}{f\oftype\vec{\base b}\arrow\base b_0}t{\base b_0}$}
\InfRuleLabel{$\crec$-I}
\BIC{$\Gtyping{\stdcrec}{\vec{\base b}\arrow\base b_0}$}
\DisplayProof
}
\\[3.5ex]
$%
\AXC{$\typing{\Gamma,v\oftype\sigma}\Delta t\tau$}
\InfRuleLabel{$\arrow$-I}
\UIC{$\GDtyping {(\lambda v.t)}{\sigma\arrow\tau}$}
\DisplayProof
$%
&
$%
\AXC{$\typing\Gamma{\Delta_0} s{\sigma\arrow\tau}$}
\AXC{$\typing\Gamma{\Delta_1} t\sigma$}
\InfRuleLabel{$\arrow$-E}
\BIC{$\typing\Gamma{\Delta_0,\Delta_1} {(st)}\tau$}
\DisplayProof
$%
\end{tabular}
\caption{$\ATR$ typing.
See the discussion for side-conditions on
\InfRuleP{$\crec$-I} and \InfRuleP{$\arrow$-E}, the definition
of $\shiftsto$, and differences
between the formalism presented here and in~\ATS.\label{fig:typing}}
\end{figure}

The intuition behind the \emph{shifts-to} relation $\shiftsto$ between
types is as follows.  Suppose $f\oftype \Nat_{\eps}\arrow\Nat_\dmnd$.
We think of $f$ as being a function that does some polynomial-time
computation to its input.  If we have an input~$x$ of type~$\Nat_{\Box\dmnd}$
then recalling the intuition behind the base types, we should be able
to assign the type~$\Nat_{\dmnd\Box\dmnd}$ to~$f(x)$.  The 
shifts-to relation allows us to shift input types in this way, with
a corresponding shift in output type.  As a concrete example,
the judgment $\typing{f\oftype\Nat_\eps\arrow\Nat_\dmnd,x\oftype\Nat_\eps}{}
{f(fx)}{\Nat_{\dmnd\Box\dmnd}}$ is derivable using \InfRuleP{Subsumption} to
coerce the type of~$f(x)$ to $\Nat_{\Box\dmnd}$ and 
\InfRuleP{Shift} to shift the type of the outer application of~$f$
to $\Nat_{\Box\dmnd}\arrow\Nat_{\dmnd\Box\dmnd}$.
The definition of~$\shiftsto$ must take into account multiple
arguments and level-$2$ types, and
it must preserve certain relationships between input and output types
(for example, shifting must ``preserve flatness''
in the sense that
if $t\oftype\sigma\arrow\tau$, $\tail(\sigma) = \tail(\tau)$, and
$\sigma\arrow\tau\shiftsto\sigma'\arrow\tau'$, then
$\tail(\sigma') = \tail(\tau')$).
Our examples in this paper (implementing lists and sorting) do not
make use of the \InfRuleP{Shift} rule, so in order to not distract the
reader from our main theme, we direct him or her to~\ATS\ for the full
definition.

\paragraph{Changes from~\ATS}
The system we present here differs from the one given
in~\ATS\ in the following ways:
\begin{enumerate}
\item \ATS\ did not restrict \InfRuleP{Shift} to have empty affine zone.
This restriction is crucial
in our discussion of plain affine recursion in 
Section~\ref{sec:plain-affine-recursion}.  Furthermore, we know of
no natural examples in which this constraint is violated.  
As \InfRuleP{Shift} provides a kind of limited polymorphism, this restriction
is similar to the restriction in ML that polymorphism is disabled
in recursive definitions (see \citet{Milner-et-al:Definition} and
\citet[Page~$338$]{Pierce:Types-and-PL}).

\item \ATS\ imposed no constraint on~$\base b_0$ 
in~\InfRuleP{$\crec$-I}.  Again, we know of no natural programs in which
this constraint is violated.

\item \ATS\ restricted~\InfRuleP{$\destr$-I}
and~\InfRuleP{$\test_a$-I} to computational types.  There was no real
need for this, as these term constructors represent
operations that are not size-increasing.

\item \ATS\ restricted~\InfRuleP{$\crec$-I} to
tail-recursion.  Of course, this is the major improvement of the
current work.

\item \ATS\ did
not allow affine variables in the argument of~\InfRuleP{$\arrow$-E}.
This is another non-trivial improvement of the current work.
\end{enumerate}

\subsection{Operational semantics}

Motivated by the approach of \citet{jones:life-wout-cons}, we define
the cost of computing a program to be the cost of a call-by-value evaluation
derivation.%
\footnote{In \ATS\ we give an abstract machine semantics based
on defunctionalized continuations; see Appendix~\ref{app:equiv-semantics}
for a proof of the equivalence between that semantics and the one we
present here.}
The evaluation relation $\evalto$ relates closures to values,
which are inductively defined as follows:%
\footnote{If one is only interested in computing, then the typing information
in the following definitions can be dropped.  However, we will address
properties of closures that arise from terms 
(specifically, bounds on the cost of evaluation) and will need to make use
of that typing information, so we include it here.}
\begin{enumerate}
\item A \emph{closure}~$\cl*{\GDtyping t\tau}\rho$ consists of a 
term $\GDtyping t\tau$ and a $(\Gamma,\Delta,t)$-environment~$\rho$.
We shall always drop reference to the explicit typing and talk
of closures~$\cl t\rho$.
\item A \emph{$(\Gamma,\Delta,t)$-environment}~$\rho$ is a finite map
from variables to extended values such that 
$\fv(t)\subseteq\dom(\Gamma,\Delta)$, $\fv(t)\subseteq\dom\rho$
and if $x\in\fv(t)$ and $(x\oftype\sigma)\in (\Gamma,\Delta)$ then
$\rho(x)$ is of type~$\sigma$.  The empty environment
is denoted~$\emptyenv$.\footnote{The only reason for including~$t$ in
this definition is so that if $t$ is a closed term with a typing
that happens to have a non-empty environment, we can still form
the closure~$\cl t\emptyenv$.}
\item A \emph{value}~$\cl z\theta$ is a closure in which~$z$ is either
a string constant, oracle, or abstraction.
\item An \emph{extended value}~$\cl z\theta$ is a closure that is 
a value or has $z = \stdcrec$ for some string constant~$a$, variables~$f$
and~$\vec v$, and term~$t$.
\end{enumerate}
For an environment~$\rho$, $\extend\rho x{\cl z\theta}$ is the
environment that is the same as~$\rho$ on variables other than~$x$,
and maps~$x$ to $\cl z\theta$.  We write
$\extend\rho{x_1,\dots,x_n}{\cl {z_1}{\theta_1},\dots,\cl{z_n}{\theta_n}}$ 
for the obvious simultaneous extension, and often abbreviate this by
$\extend\rho{\vec x}{\vec{\cl z\theta}}$ or 
$\extend\rho{x_i}{\cl{z_i}{\theta_i}}$, where in the latter $i$ has a range
that should be clear from context.  We will also occasionally
write $\extend\rho{x_{i..j}}{\cl{z_{i..j}}{\theta_{i..j}}}$ for
$\extend\rho{x_i,\dots,x_j}{\cl{z_i}{\theta_i},\dots,\cl{z_j}{\theta_j}}$.

The evaluation relation $\cl t\rho\evalto\cl z\theta$ is defined
in Figure~\ref{fig:eval}.  It is a fairly standard call-by-value
operational semantics; we just make a few points about some of the rules:
\begin{itemize}
\item Because environments may assign \keyw{crec} terms to variables,
we cannot assume that $\rho(x)$ is a value in \InfRuleP{Env}.  However,
we note that $\rho(x)\evalto\cl z\theta$ is an instance of either
the \InfRuleP{Val} or \InfRuleP{$\crec$} axioms.
\item In the \InfRuleP{$\crec$} rule, ``$\lh a\leq\lh{v_1}$'' is shorthand
for $\down (\cons_0 a)(\cons_0 v_1)$.
\item In the \InfRuleP{$\down_i$} rules, $a_s$ and $a_t$ are string
constants, so the length comparison makes sense.  Our cost model will take
into account the actual cost of the length comparison.
\item 
Recalling that oracles name type-$1$ functions and that the only
type-$0$ values are string constants, the evaluation rules~$O_0$ and~$O_1$
say to treat multiple-argument oracles as though they are in curried
form, returning the curried oracle result until all arguments have
been provided.
\end{itemize}
%%%
%%% Evaluation relation
%%%
\begin{figure}[t]
\begin{gather*}
%% Axioms
\AXC{}
\RightLabel{($\cl z\theta$ a value)}
\LeftLabel{\InfRule{Val}}
\UIC{$\cl z\theta\evalto\cl z\theta$}
\DisplayProof
\\
\AXC{}
\LeftLabel{\InfRule{$\crec$}}
\UIC{$\cl*{\stdcrec}\rho\evalto\cl*{\lambda\vec v.\cond{\lh{a}\leq\lh{v_1}}{t}{\eps}}{\extend\rho f{\crec(\mathbf{0}a)(\afflambda f.\lambda\vec v.t)}}$}
\DisplayProof
\\
%% Environment
\AXC{$\rho(x)\evalto\cl z\theta$}
\InfRuleLabel{Env}
\UIC{$\cl x\rho\evalto\cl z\theta$}
\DisplayProof
\\
%%% cons and destr
\AXC{$\cl s\rho\evalto\cl a\theta$}
\LeftLabel{\InfRule{$\cons_b$}}
\RightLabel{($\mathbf b=\bz,\bone$)}
\UIC{$\cl*{\cons_{\mathbf b} s}\rho\evalto\cl*{\mathbf{b}a}\theta$}
\DisplayProof
\qquad
\AXC{$\cl s\rho\evalto\cl\eps\theta$}
\LeftLabel{\InfRule{$\destr_0$}}
\UIC{$\cl*{\destr s}\rho\evalto\cl\eps\theta$}
\DisplayProof
\qquad
\AXC{$\cl s\rho\evalto\cl*{\mathbf{b}a}\theta$}
\LeftLabel{\InfRule{$\destr_1$}}
\UIC{$\cl*{\destr s}\rho\evalto\cl a\theta$}
\RightLabel{($\mathbf b=\bz,\bone$)}
\DisplayProof
\\
%% test
\AXC{$\cl s\rho\evalto\cl a\theta$}
\RightLabel{($a\not=\mathbf ba'$ any~$a'$)}
\LeftLabel{\InfRule{$\test_0$}}
\UIC{$\cl*{\test_{\mathbf b} s}\rho\evalto\cl\eps\emptyenv$}
\DisplayProof
\qquad
\AXC{$\cl s\rho\evalto \cl*{\mathbf ba}\theta$}
\LeftLabel{\InfRule{$\test_1$}}
\UIC{$\cl*{\test_{\mathbf b} s}\rho\evalto\cl{\mathbf{0}}\emptyenv$}
\DisplayProof
\\
%% down
\AXC{$\cl s\rho\evalto\cl{a_s}\theta_s$}
\AXC{$\cl t\rho\evalto\cl{a_t}\theta_t$}
\AXC{$\lh{a_s}\leq\lh{a_t}$}
%\InfRuleLabel{$\down_0$}
\LeftLabel{\InfRule{$\down_0$}}
\TIC{$\cl*{\down s\,t}\rho\evalto \cl{a_s}\theta_s$}
\DisplayProof
\\
\AXC{$\cl s\rho\evalto\cl{a_s}\theta_s$}
\AXC{$\cl t\rho\evalto\cl{a_t}\theta_t$}
\AXC{$\lh{a_s}>\lh{a_t}$}
\LeftLabel{\InfRule{$\down_1$}}
\TIC{$\cl*{\down s\,t}\rho\evalto \cl{\eps}\emptyenv$}
\DisplayProof
\\
%% Conditional
\AXC{$\cl s\rho\evalto\cl{a}\theta'$}
\AXC{$\cl{t_0}\rho\evalto\cl z\theta$}
\RightLabel{($a\not=\eps$)}
\LeftLabel{\InfRule{$\keyw{if}_0$}}
\BIC{$\cl*{\cond{s}{t_0}{t_1}}\rho\evalto \cl z\theta$}
\DisplayProof
\qquad
\AXC{$\cl s\rho\evalto\cl\eps\theta'$}
\AXC{$\cl{t_1}\rho\evalto\cl z\theta$}
\LeftLabel{\InfRule{$\keyw{if}_1$}}
\BIC{$\cl*{\cond{s}{t_0}{t_1}}\rho\evalto \cl z\theta$}
\DisplayProof
\\
%% Application
\AXC{$\cl s\rho\evalto\cl*{\lambda x.s'}\theta'$}
\AXC{$\cl t\rho\evalto\cl {z''}{\theta''}$}
\AXC{$\cl{s'}{\extend{\theta'} x{\cl {z''}{\theta''}}}\evalto \cl z\theta$}
\LeftLabel{\InfRule{App}}
\TIC{$\cl*{st}\rho\evalto \cl z\theta$}
\DisplayProof
\\
%% Oracles
\AXC{$\cl s\rho\evalto\cl {\alpha^{\base b_1\arrow\base b}}\theta'$}
\AXC{$\cl t\rho\evalto\cl a\theta$}
\AXC{$\alpha(a) = a'$}
\InfRuleLabel{$O_0$}
%\RightLabel{($\alpha$ unary)}
\TIC{$\cl*{st}\rho\evalto \cl a'\emptyenv$}
\DisplayProof
\\
\AXC{$\cl s\rho\evalto\cl {\alpha^{(\base b_1,\dots,\base b_k)\arrow\base b}}\theta'$}
\AXC{$\cl t\rho\evalto\cl a\theta$}
\AXC{$\alpha(a) = \alpha'$}
\InfRuleLabel{$O_1$}
\RightLabel{($k\geq 2$)}
\TIC{$\cl*{st}\rho\evalto \cl {\alpha'}\emptyenv$}
\DisplayProof
\end{gather*}
\caption{$\ATR$ evaluation.  Note that in the $O_i$ rules
$a$ is necessarily a string constant, hence $\theta$ is 
irrelevant.\label{fig:eval}}
\end{figure}

The cost of a derivation is the sum of the costs of the rules.  All
rules have cost~$1$ except:
% Can't follow an \item by an \hbox?!?
\begin{itemize}
\item\hspace{0pt}\InfRuleP{Env}:  
if $z$ is a string constant this rule has cost $1\bmax\lh{z}$;
otherwise if $z$ is an abstraction or oracle, this rule has cost~$1$.
This reflects a length-cost model of accessing the environment, where
string constants are copied into memory bit-by-bit, but higher-type
values are simply stored in memory as references.
\item\hspace{0pt}\InfRuleP{$\keyw{down}_i$}:  the cost of this rule is $2\lh{a_t}+1$.
This reflects the cost of comparing $a_s$ and $a_t$ bit-by-bit to determine
which is longer.
\item\hspace{0pt}\InfRuleP{$O_0$}:  the cost of this rule is $1\bmax\lh{a'}$, similar
to accessing a base-type value in the environment.
\item\hspace{0pt}\InfRuleP{$O_1$}:  the cost of this rule is $1$, similar 
to accessing a higher-type value in the environment.
\end{itemize}

\begin{defn}
\label{defn:cost}
$\cost(\cl t\rho)$ is defined to be the cost of the evaluation derivation
of~$\cl t\rho$.
We write $\cl t\rho\evalto_n\cl z\theta$ to indicate that
$\cl t\rho\evalto\cl z\theta$ and $\cost(\cl t\rho)\leq n$.
\end{defn}

\emph{A priori} $\cost(\cl t\rho)$ may be infinite, as there may not be
an evaluation derivation of $\cl t\rho$.  Intuitively the problem
may be that the ``clock'' $\lh{v_1}$ in the
\InfRuleP{$\crec$} rule may be increased during the recursive call,
thus leading to a non-terminating recursion.  The main work of this
paper to show that $\cost(\cl t\rho)$ is always finite and in fact
second-order polynomially bounded.

\section{Programming in $\ATR$}
\label{sec:programming}

To illustrate $\ATR$ programming we 
give a data-type implementation of lists of binary strings and then
present versions of insertion- and selection-sort using this implementation.
These programs are fairly close to straightforward ML for these algorithms,
with a few crucial differences discussed below.  Also,
lists and both sorts nicely highlight various forms
of affine recursion that we will need to treat in our analysis
of the complexity properties of $\ATR$ programs.

\sloppypar In these programs
we use the ML
notation \lstinline[]!fn x$\; \Rightarrow\dotsc$!
for $\lambda$-abstraction.  Also
\lstinline[]!let val x=s in t end! abbreviates
\lstinline[]!(fn x$\; \Rightarrow\;$ t)s! and
\lstinline[]!letrec f=s in t end! abbreviates
$\pgm{t}[f\mapsto\crec\eps(\afflambda f.s)]$.

We implement lists of binary words as 
concatenated self-delimiting strings.  Specifically,
we code the word $w=b_0\dots b_{k-1}$ 
as $s(w) = 1b_01b_1\dots1b_{k-1}0$ and the list
$\langle w_0,\dots,w_{k-1}\rangle$
as $s(w_0)\cat\dots\cat s(w_{k-1})$, where $\cat$ is the concatenation
operation.  Code for the basic list operations is given in
Figure~\ref{fig:list-ops}.
Note that the $\pgm{cons}$, $\pgm{head}$, and $\pgm{tail}$
programs all use cons-tail recursion---that is, the application of the
recursively-defined function is followed by some number of basic
operations. Insertion-sort is expressed in
essentially its standard form, as in Figure~\ref{fig:insert-sort}.
This implementation requires another form of recursion, in
which the complete application of the recursively-defined
function appears in an argument to some operator.
Selection-sort (Figure~\ref{fig:sel-sort})
requires yet another form of recursion
in which the complete application of the recursively-defined function
appears in the body of a $\keyw{let}$-expression.
All of these recursion schemes are special cases of what we
call \emph{plain affine recursion}, which we discuss in
Section~\ref{sec:plain-affine-recursion}.
\begin{figure}[tb]
%\lstinputlisting[linerange=10-15]{list-ops.atr}
\begin{lstlisting}
val nil = $\eps$ : $\Nat_{\eps}$

val cons : $\Nat_{\eps}\arrow\Nat_{\dmnd}\arrow\Nat_{\dmnd}$ = 
    fn x xs $\Rightarrow$ letrec enc : $\Nat_{\eps}\arrow\Nat_{\dmnd}\arrow\Nat_{\dmnd}$ =
        fn b y $\Rightarrow$ if y then if $\test_0$(y) then $\cons_1$($\cons_0$(enc b (d y)))
                          else $\cons_1$($\cons_1$(enc b (d y)))
                  else $\cons_0$(xs)
    in enc x x end

val head : $\Nat_{\eps}\arrow\Nat_{\eps}$ =
    fn xs $\Rightarrow$ letrec dec : $\Nat_{\eps}\arrow\Nat_{\dmnd}\arrow\Nat_{\dmnd}$ =
        fn b ys $\Rightarrow$ if $\test_1$(ys) then
                      if $\test_0$(d ys) then $\cons_0$(dec b (d(d(ys)))) else $\cons_1$(dec b (d(d(ys))))
                   else $\eps$
    in down (dec xs xs)(xs) end

val tail : $\Nat_{\eps}\arrow\Nat_{\eps}$ =
    fn xs $\Rightarrow$ letrec strip : $\Nat_{\eps}\arrow\Nat_{\eps}\arrow\Nat_{\eps}$ =
        fn b ys $\Rightarrow$ if $\test_1$(ys) then strip b d(d(ys)) else d(ys)
    in strip xs xs end
\end{lstlisting}
\caption{The basic list operations in~$\ATR$.\label{fig:list-ops}}
\end{figure}
\begin{figure}[tb]
\begin{lstlisting}
val insert : $\Nat_{\eps}\arrow\Nat_{\eps}\arrow\Nat_{\dmnd}$ =
    fn x xs $\Rightarrow$ letrec ins : $\Nat_{\eps}\arrow\Nat_{\eps}\arrow\Nat_{\dmnd}$ =
        fn b ys $\Rightarrow$ if ys then
                       if leq x head(ys) then cons x ys
                       else cons (head ys) (ins b (tail ys))
                   else cons x nil
    in ins xs xs end

val ins_sort : $\Nat_{\eps}\arrow\Nat_{\dmnd}$ =
    fn xs $\Rightarrow$ letrec isort : $\Nat_{\eps}\arrow\Nat_{\eps}\arrow\Nat_{\dmnd}$ =
        fn b ys = if ys then insert (head ys) (down (isort b (tail ys)) ys) else $\eps$
    in isort xs xs end
\end{lstlisting}
\caption{Insertion-sort in~$\ATR$.
The function \lstinline!leq! tests two integers written
in binary for inequality; we leave its full definition as an exercise for
the reader.
\label{fig:insert-sort}}
\end{figure}
\begin{figure}[tb]
\begin{lstlisting}
val swap : $\Nat_\eps\arrow\Nat_\eps\arrow\Nat_\dmnd$ =
    fn x xs $\Rightarrow$ if leq x (head xs) then cons x xs
               else cons (head xs) (cons x (tail xs))

val select : $\Nat_\eps\arrow\Nat_\eps$ =
    fn xs $\Rightarrow$ letrec sel $\oftype \Nat_\eps\arrow\Nat_\eps\arrow\Nat_\eps$ =
        fn b ys $\Rightarrow$ if tail(ys) then down (swap (head ys) (sel b (tail ys))) ys 
                   else ys
    in sel xs xs end

val sel_sort : $\Nat_\eps\arrow\Nat_\dmnd$ =
    fn xs $\Rightarrow$ letrec ssort $\oftype \Nat_\eps\arrow\Nat_\eps\arrow\Nat_\dmnd$ =
        fn b ys $\Rightarrow$ let val m = select ys in cons (head m) (ssort b (tail m)) end
    in ssort xs xs end
\end{lstlisting}
\caption{Selection-sort in~$\ATR$.\label{fig:sel-sort}.}
\end{figure}

Our  $\pgm{head}$ and $\pgm{ins\_sort}$
programs  use the $\down$ operator to coerce the type~$\Nat_{\dmnd}$
to~$\Nat_{\eps}$.  Roughly, $\down$ is used in places
where our type-system is not clever enough to prove that the result of a
recursion is of size no larger than one of the recursion's initial
arguments; the burden of supplying these proofs is shifted off to the
correctness argument for the recursion.   A cleverer type system (say,
along the lines of Hofmann's \citep{Hofmann:ic03}) 
could obviate many of these $\down$'s,
but at the price of more complex syntax (i.e., typing), semantics (of
values and of time-complexities), and, perhaps, pragmatics (i.e.,
programming).  Our use of $\down$ gives us a more primitive (and
intensional) system than found in pure implicit complexity,%
\footnote{Leivant's \emph{recursion under a
high-tier bound}~\citep[\S3.1]{Leivant:Ram-rec-I} implements a similar idea.}
but it
also gives us a less cluttered setting to work out the basics of
complexity-theoretic compositional semantics---the focus of the rest
of the paper.  Also, in practice the proofs that the uses of $\down$
forces into the correctness argument are for the most part obvious,
and thus not a large burden on the programmer.

\section{Time-complexity semantics and soundness for non-recursive terms}
\label{sec:t-c-survey}

The key fact we want to establish about~$\ATR$ and its operational
semantics is that the cost of evaluating a term to a value is, in an
appropriate sense, polynomially bounded.  This section sets up the
framework for proving this and establishes the result for non-recursive
terms.

The key technical notion is that of \emph{bounding} a closure~$\cl t\rho$
by a \emph{time-complexity}, which
provides upper bounds on both the \emph{cost} of evaluating 
$\cl t\rho$ to a value~$\cl z\theta$ as well as the 
\emph{potential} cost of using~$\cl z\theta$.  The potential of
a base-type closure is just its (denotation's) length, whereas the
potential of a function~$f$ is itself a function that maps potentials~$p$ to
the time complexity of evaluating~$f$ on arguments of potential~$p$
(more on this later---we give precise definitions in
Section~\ref{sec:time-complexity-semantics}).
The bounding relation gives a
\emph{time-complexity semantics} for~$\ATR$-terms; a \emph{soundness theorem}
asserts the existence of a bounding time-complexity for every $\ATR$~term.
In this paper, our soundness theorems also assert that the bounding
time-complexities are \emph{safe} (Definition~\ref{defn:safety}),
which in particular implies type-2 polynomial size and cost bounds
for the closure.  We thereby encapsulate the Soundness,
polynomial-size-boundedness, and polynomial-time-boundedness theorems of~\ATS\
(the \emph{value semantics} for the meaning of~$\ATR$ terms
and corresponding soundness theorem are essentially
unchanged).

\subsection{Time-complexity semantics}
\label{sec:time-complexity-semantics}

Our prior discussion of~$\ATR$ types and terms situated their
semantics in the realm of values---i.e., $\bz$-$\bone$-strings,
functions over strings, functionals over functions over strings, etc.
To work with time-complexities and potentials we introduce a new
type system and new semantic realm for bounds.  We will connect
the realms of values and bounds in Definition~\ref{defn:bounding-relation}
where we introduce bounding relations.

We start by defining \emph{cost}, \emph{potential},
and \emph{time-complexity} types, all of which are elements of the
simple product type structure over the \emph{time-complexity base types}
$\set{\Tally}\union\setst{\Tally_L}{\text{$L$ is a label}}$.
The intended interpretation of these base types is the unary numerals
and of product types the usual cartesian product.
The arrow types are interpreted as the pointwise monotone non-decreasing
functions and are further ``pruned'' analogously
to the well-tempered semantics for~$\ATR$ (see the discussion
following Definition~\ref{defn:predicative-etc})---for more details
see Section~$12$ of~\ATS\ and in particular Definition~$49$.

We define a subtype relation on base types by
$\Tally_L\subtype\Tally_{L'}$ if $L\leq L'$ and $\Tally_L\subtype\Tally$
for all~$L$, and extend it to product and function types in the standard
way.  
The only \emph{cost type} is $\Tally$.  For each $\ATR$-type~$\sigma$ we
define the 
\emph{time-complexity type}~$\tcden\sigma$ and
\emph{potential type}~$\potden\sigma$ by
\[
\tcden\tau = \Tally\cross\potden\tau
\qquad
\potden{\Nat_L} = \Tally_L
\qquad
\potden{\sigma\arrow\tau} = \potden\sigma\arrow\tcden\tau.
\]
We denote the left- and right-projections
on~$\tcden\tau$ by $\tccost(\cdot)$ and $\pot(\cdot)$, respectively.  Define 
$\tail(\tcden\tau) = \potden{\tail(\tau)}$.  Extend
the notions of predicative, impredicative, etc. from
Definition~\ref{defn:predicative-etc} to time-complexity and
potential types in the obvious way.
We note that $\tcden\sigma\subtype\tcden\tau$ iff $\sigma\subtype\tau$.
We define $\tcden\sigma\shiftsto\tcden\tau$ if $\sigma\shiftsto\tau$ and
$\potden\sigma\shiftsto\potden\tau$ if $\sigma\shiftsto\tau$.

We will need to describe objects in the time-complexity types and
introduce a small formalism to do so.
We will only consider terms of cost, potential, and time-complexity type.
We use a fresh set of variables that we call \emph{time-complexity
variables} and for each $\ATR$ oracle symbol $\alpha^\sigma$ we
have a \emph{time-complexity oracle symbol} $\alpha^{\tcden\sigma}$.
Define a \emph{time-complexity context} 
to be a finite map from t.c.\ variables
to cost and potential types.\footnote{For obvious reasons, we shall 
start abbreviating ``time-complexity'' as ``t.c.''}
For a t.c.\ context~$\Sigma$, a \emph{$\Sigma$-environment} is a
finite map from $\dom\Sigma$ to the interpretation of the time-complexity
types that respects the type $\Sigma$ assigns to each variable;
we denote the set of $\Sigma$-environments by $\Env{\Sigma}$.
We use the same extension notation for t.c.\ environments as for
term environments.
We extend $\tcden\cdot$
to $\ATR$-type contexts by introducing t.c.\ variables~$x_c$ and~$x_p$ for
each $\ATR$-variable~$x$ and setting
$\tcden{\Gamma;\Delta} = \union_{(x\oftype\sigma)\in(\Gamma;\Delta)}\set{x_c\oftype\Tally,x_p\oftype\potden\sigma}$.
A \emph{time-complexity denotation} of 
t.c.\ type~$\gamma$ w.r.t.\ a t.c.\ environment $\Sigma$ is a function
$X:\Env{\Sigma}\to\gamma$.
The projections $\tccost$ and $\pot$ extend to t.c.\ denotations 
as $\tccost(X) = \varrho\mapsto\tccost(X\varrho)$ and
$\pot(X) = \varrho\mapsto\pot(X\varrho)$.
We now come to the main technical notion, that of bounding
a term by a t.c.\ denontation.

\begin{defn} ~
\label{defn:bounding-relation}
\begin{enumerate}
\item 
Suppose $\cl t\rho$ is a closure and $\cl z\theta$ a value, both
of type~$\tau$; $\chi$ a time-complexity of type~$\tcden\tau$; and
$q$ a potential of type~$\potden\tau$.  Define the
\emph{bounding relations}
$\cl t\rho\apprby^\tau\chi$ and
$\cl z\theta\apprbypot^\tau q$ as follows:\footnote{We will drop the superscript
when it is clear from context.}
\begin{enumerate}
\item $\cl t\rho\apprby^\tau\chi$ if
    $\cl t\rho\evalto_{\tccost(\chi)}\cl z\theta$ and
    $\cl z\theta\apprbypot^\tau\pot(\chi)$
	(recall that the subscript on $\evalto$ indicates an upper bound on the
	cost of the evaluation derivation).
\item $\cl z\theta\apprbypot^{\base b} q$ if $\lh z\leq q$.
\item $\cl*{\lambda v.t}{\theta}\apprbypot^{\sigma\arrow\tau}q$ when
    for all values~$\cl z\eta$ and all potentials~$p$,
	if $\cl z\eta\apprbypot^{\sigma} p$,
    then $\cl{t}{\extend\theta v{\cl z\eta}}\apprby^\tau q(p)$.
\item $\cl \alpha\theta\apprbypot^{\sigma\arrow\tau}q$ when for
    all values~$\cl z\eta$, if $\cl z\eta\apprbypot^{\sigma} p$, then
    $\cl*{\alpha(\cl z\eta)}\emptyenv\apprby^\tau q(p)$.
\end{enumerate}
\item For $\rho\in\Env{(\Gamma;\Delta)}$ and 
$\varrho\in\Env{\tcden{\Gamma;\Delta}}$, we
write $\rho\apprby\varrho$ if for all $v\in\dom\rho$ we have that
$\cl v\rho\apprby(\varrho(v_c),\varrho(v_p))$.
\item For an $\ATR$-term $\GDtyping t\tau$ and a time-complexity
denotation $X$ of type~$\tcden\tau$ w.r.t.~$\tcden{\Gamma;\Delta}$,
we say $t\apprby X$ if for all~$\rho\in\Env{(\Gamma;\Delta)}$ and
$\varrho\in\Env{\tcden{\Gamma;\Delta}}$ such that
$\rho\apprby\varrho$ we have that $\cl t\rho\apprby X\varrho$.
\end{enumerate}
\end{defn}

We define second-order polynomial expressions of cost, potential,
and time-complexity types using the operations
$+$, $*$, and $\bmax$ (plus, times, and binary maximum); 
the typing rules are given
in Figure~\ref{fig:poly-typing}.
Of course, a polynomial
$\tctyping\Sigma p\gamma$ corresponds to a t.c.\ denotation
of type~$\gamma$ w.r.t.\ $\Sigma$ in the obvious way.  We shall
frequently write $p_p$ for $\pot(p)$.  Our primary interest is
in constructing a bounding
t.c.\ polynomial $\tctyping{\tcden{\Gamma;\Delta}}p{\tcden\tau}$
for each term $\GDtyping t\tau$.  Rather than writing
$p = \dotsb(x_c,x_p)\dotsb$ each $x\in\dom(\Gamma\union\Delta)$, we shall
just write $p = \dotsb x\dotsb$.
\begin{figure}[t]
\begin{gather*}
\AXC{}
\UIC{$\Stctyping \eps {\Tally_\eps}$}
\DisplayProof
\quad
\AXC{}
\UIC{$\Stctyping {\mathbf{0}^n} {\Tally_\dmnd}$}
\DisplayProof
\quad
\AXC{}
\UIC{$\Stctyping {\alpha^{\tcden\sigma}}{\tcden\sigma}$}
\DisplayProof
\\
\AXC{}
\UIC{$\tctyping {\Sigma,x\oftype\gamma} {x} \gamma$}
\DisplayProof
\\
\AXC{$\Stctyping p \gamma$}
\RightLabel{($\gamma\shiftsto\gamma'$)}
\UIC{$\Stctyping p {\gamma'}$}
\DisplayProof
\quad
\AXC{$\Stctyping p \gamma$}
\RightLabel{($\gamma\subtype\gamma'$)}
\UIC{$\Stctyping p {\gamma'}$}
\DisplayProof
\\
\AXC{$\Stctyping p {\base b}$}
\AXC{$\Stctyping q {\base b}$}
\BIC{$\Stctyping {p\bullet q}{\base b}$}
\DisplayProof
\quad
\AXC{$\Stctyping p {\base b}$}
\AXC{$\Stctyping q {\base b}$}
\BIC{$\Stctyping {p\bmax q}\base b$}
\DisplayProof
\\
\AXC{$\tctyping{\Sigma, x\oftype\potden\sigma}{p}{\tcden\tau}$}
\UIC{$\tctyping\Sigma {\lambda x.p} {\potden{\sigma\arrow\tau}}$}
\DisplayProof
\quad
\AXC{$\Stctyping p {\potden{\sigma\arrow\tau}}$}
\AXC{$\Stctyping q {\potden\sigma}$}
\BIC{$\Stctyping {pq} {\tcden\tau}$}
\DisplayProof
\\
\AXC{$\Stctyping p \Tally$}
\AXC{$\Stctyping q {\potden\tau}$}
\BIC{$\Stctyping {(p,q)} {\tcden\tau}$}
\DisplayProof
\quad
\AXC{$\Stctyping p {\tcden\tau}$}
\UIC{$\Stctyping {\tccost(p)} {\Tally}$}
\DisplayProof
\quad
\AXC{$\Stctyping p {\tcden\tau}$}
\UIC{$\Stctyping {\pot(p)} {\potden\tau}$}
\DisplayProof
\end{gather*}
\caption{Typing rules for time-complexity polynomials.
The type~$\base b$ is a t.c.\ base type,
$\gamma$ and $\gamma'$ are any t.c.\ or potential types,
and $\sigma$ and $\tau$ are any $\ATR$-types.
The operation~$\bullet$ is $+$ or $*$ and in this rule
$\base b$ is either $\Tally$ or $\Tally_{\dmnd_k}$ for some~$k$.
\label{fig:poly-typing}}
\end{figure}

\begin{defn}
\label{defn:shadowed}
Suppose $\Stctyping p\gamma$ is a t.c.\ polynomial and $s$ is a subterm
occurrence of~$p$.  We say that $s$ is \emph{shadowed} if (1)~$s$ occurs in a
context $ts$ where the occurence of~$t$ has impredicative 
type~$\sigma\arrow\tau$ with $\tail(\tau)\strsubtype\tail(\sigma)$,
or (2)~the occurrence of~$s$ appears properly within another shadowed
subterm occurrence.
\end{defn}

\begin{defn}
\label{defn:safety}
Let $\gamma$ be a potential type, $\base b$ a time-complexity base type, 
$p$ a potential polynomial, and suppose $\Stctyping p\gamma$.
\begin{enumerate}
\item $p$ is \emph{$\base b$-strict}
w.r.t.~$\Sigma$ when $\tail(\gamma)\subtype\base b$
and every unshadowed
free-variable occurrence in~$p$ has a type
with tail $\strictsubtype\base b$.
\item $p$ is \emph{$\base b$-chary}
w.r.t.~$\Sigma$ when $\gamma=\base b$ and
$p = p_1\bmax\dots\bmax p_m$ with $m\geq 0$ where 
$p_i = (vq_1\dots q_k)$ with $v$ a variable or oracle symbol
and each $q_j$ $\base b$-strict w.r.t.~$\Sigma$.
As special cases we get $p=0$ ($m=0$) and $p=v$ for $v$ a base-type
potential variable ($m=1$ and $k=0$).
\item $p$ is \emph{$\base b$-safe} w.r.t.~$\Sigma$ if:
    \begin{enumerate}
    \item $\gamma$ is a base type and $p = q\pmjb r$ where
        $q$ is $\base b$-strict and $r$ is $\base b$-chary,
        $\pmjb = \bmax$ if $\base b$ is oracular, and
        $\pmjb = +$ if $\base b$ is computational.
    \item $\gamma=\potden{\sigma\arrow\tau}$ and $\pot(pv)$ is 
        $\base b$-safe
        w.r.t.~$\Sigma,v\oftype\potden\sigma$.
    \end{enumerate}
\item A t.c.\ polynomial $\Stctyping q{\tcden\tau}$ 
is \emph{$\base b$-safe} if $\pot(q)$ is.
\item
A t.c.\ denotation $X$ of type~$\tcden\tau$
w.r.t.~$\Sigma$ is \emph{$\base b$-safe} if there is
a $\base b$-safe t.c.\ polynomial $\Stctyping p{\tcden\tau}$
such that $X\leq p$.\footnote{Remember that this inequality is with
respect to the well-tempered semantics discussed at the beginning
of this section.}
$X$ is \emph{safe} if $X$ is $\tail(\tcden\tau)$-safe.%
\end{enumerate}
\end{defn}

For full details and basic properties of safety, see \ATS\ Section~$8$.
Here we just give a couple of example propositions to get a feel
for how to manipulate safe polynomials.

\begin{prop}
\label{no-cost-vars-in-pot-polys}
If $\tctyping{\Sigma,x\oftype\Tally} p{\Tally_L}$ 
is a $\Tally_L$-safe polynomial, 
then every occurrence of $x$ in $p$ is shadowed.
\end{prop}
\begin{proof}
Set $\base b = \Tally_L$.
We have that $p = q\pmjb r$ where $q$ is $\base b$-strict and $r$ is
$\base b$-chary.  Since $q$ is $\base b$-strict and
$\Tally_L\subtype\Tally$, any occurrence of~$x$ must be shadowed in~$q$.
The polynomial $r$ cannot have the form~$\dotsb\bmax x\bmax\dotsb$ because
this latter expression can only have type~$\Tally$.  Thus any occurrence
of~$x$ in~$r$ must occur in some~$\base b$-strict polynomial, and the
argument just given tells us that any such occurrence must be shadowed.
\end{proof}

Under the well-tempered semantics, shadowed subterms do not contribute
to the value of a polynomial.  Thus we can w.l.o.g.\ assume that any
safe potential polynomial contains only variables of potential type by
replacing every occurrence of every variable of type~$\Tally$ with~$\eps$.

\begin{prop}
\label{bound-max-of-poly}
If $p$ and $p'$ are $\base b$-safe potential polynomials,
then there is a $\base b$-safe potential polynomial~$p^*$ such
that $p\bmax p'\leq p^*$.
\end{prop}
\begin{proof}
If $\base b$ is computational, then $p = q+r$ and $p' = q'+r'$ where
$q$ and $q'$ are $\base b$-strict and $r$ and $r'$ are $\base b$-chary.
Thus $p+p' = (q+r)\bmax (q'+r') \leq q+q'+(r\bmax r')$ is $\base b$-safe.
Similarly, if $\base b$ is oracular, then
$p+p' = (q\bmax r)\bmax (q\bmax r') = (q\bmax q')\bmax(r\bmax r')$.
\end{proof}

\subsection{Soundness for non-recursive terms}
The Soundness Theorem asserts that every 
term is bounded by a safe t.c.\ denotation; in particular,
the potential component is bounded by a safe type-$2$ polynomial
(we shall also be able to conclude that the cost component
is bounded by a type-2 polynomial in the lengths of~$t$'s free variables).
At base type, the statement about the potential
corresponds to the ``poly-max'' bounds that can be computed
for Bellantoni-Cook and Leivant-style tiered functions
(e.g., \citep[Lemma 4.1]{Bellantoni-Cook:Recursion-theoretic-char}).
The bulk of the work is in handling $\crec$ terms.  To ease the presentation,
we first extract out the main claim for $\ATRm$, the sub-system of~$\ATR$
that does not include~$\crec$.  Although we could prove a version of
the Soundness Theorem directly for $\ATRm$ by structural induction on terms, 
we state instead a slightly
more general proposition from which the Soundness Theorem follows directly.
The reason is that when analyzing~$\crec$ terms
we will frequently need to construct bounding
t.c.\ denotations for terms~$t$ given assumptions about bounding
t.c.\ denotations for the subterms of~$t$.  Thus we need to extract
out what is really just the induction step of the proof of 
the $\ATRm$ Soundness Theorem into its own lemma
(Lemma~\ref{atr-minus-soundness-istep}).

% Pisses me off - this is supposed to work immediately after the \subsection.
\suppressfloats

Figure~\ref{fig:atr-minus-tc-ops} gives a number of operations on time
complexity denotations that correspond to the~$\ATRm$ term-forming operations
other than application and abstraction.  In that figure and the following,
we use the notation $\llambda x.\cdots$ to denote the (semantic)
map $x\mapsto\cdots$.
\begin{figure}[t]
\begin{align*}
\tcdcons&:X\mapsto \llambda\varrho\bigl(1+\tccost(X\varrho),\;1+\pot(X\varrho)\bigr) \\
\tcddest&:X\mapsto \llambda\varrho\bigl(1+\tccost(X\varrho),\;\pot(X\varrho)\bigr) \\
\tcdtest&:X\mapsto \llambda\varrho\bigl(1+\tccost(X\varrho),\;1\bigr) \\
\tcdcond&:(X,Y,Z)\mapsto\llambda\varrho\bigl(1+\tccost(X\varrho)+(\tccost(Y\varrho)\bmax\tccost(Z\varrho)),\\
 &\phantom{:(X,Y,Z)\mapsto\llambda\varrho\bigl(}\qquad\pot(Y\varrho)\bmax\pot(Z\varrho)\bigr) \\
\tcddown&:(X,Y)\mapsto \llambda\varrho\bigl(1+\tccost(X\varrho)+\tccost(Y\varrho)+2\pot(Y\varrho),\;\pot(Y\varrho)\bigr)
\end{align*}
\caption{Operations on time-complexity denotations of base type.\label{fig:atr-minus-tc-ops}}
\end{figure}
For application and abstraction, we make the following definitions:

\begin{defn}~
\begin{enumerate}
\item For a potential~$p$, if $p$ is of base type, $\val p = (1\bmax p,p)$;
if $p$ is of higher type, then $\val p = (1, p)$.\footnote{Notice that
$\val(p)$ is a time-complexity that bounds a value with potential~$p$.}  For
a t.c.\ environment~$\varrho$ and $\ATR$ variable~$v$ we
write $\extend\varrho v{\chi}$ for
$\extend\varrho{v_c,v_p}{\tccost(\chi),\pot(\chi)}$.
\item If $Y$ is a t.c.\ denotation of type~$\tcden\tau$
w.r.t.~$\Sigma,\tcden{v\oftype\sigma}$, then
\[
  \llambda_\star v.Y \eqdef \llambda\varrho\bigl(1,\; \llambda v_p.Y(\extend\varrho v {\val v_p})\bigr)
\]
is a t.c.\ denotation of type~$\tcden{\sigma\arrow\tau}$
w.r.t.~$\Sigma$.
\item If $X$ and $Y$ are t.c.\ denotations of type $\tcden{\sigma\arrow\tau}$
and $\tcden\tau$ w.r.t.\ $\Sigma$, then
\[
  X\star Y \eqdef \llambda\varrho\bigl(\tccost(X\varrho)+\tccost(Y\varrho)+\tccost(\chi)+1,\, \pot(\chi)\bigr)
\]
is a t.c.\ denotation of type~$\tcden\tau$ w.r.t.\ $\Sigma$, where
$\chi = \pot(X\varrho)(\pot(Y\varrho))$.
For $\vec Y = Y_1,\dots, Y_k$ we write
$X\star\vec Y$ for $X\star Y_1\star\dots\star Y_k =
((X\star Y_1)\star\dots)\star Y_k$.
\end{enumerate}
\end{defn}

The key lemma is the following; the apparent complexity is solely
due to our embedding of what would normally be an induction hypothesis
into the statement of the lemma itself.

\begin{lem}~
\label{atr-minus-soundness-istep}
\begin{enumerate}
\item \label{item:easy}
Suppose $\GDtyping r{\base b}$, $\GDtyping s{\base b'}$, 
$\GDtyping t{\base b'}$
and that
$X$, $Y$, and $Z$ are t.c.\ denotations of types $\tcden{\base b}$,
$\tcden{\base b'}$, and $\tcden{\base b'}$ 
w.r.t. $\tcden{\Gamma;\Delta}$ respectively
such that $r\bddby X$, $s\bddby Y$, and $t\bddby Z$.  Then:
\begin{enumerate}
\item $\cons_a r\apprby\tcdcons X$,
$\destr r\apprby \tcddest X$, and $\test_a r\apprby \tcdtest X$.
\item $\cond rst\apprby \tcdcond(X,Y,Z)$.
\item $\down r\,s\apprby \tcddown(X,Y)$.
\end{enumerate}
\item \label{item:abs}
If $\typing{\Gamma,v\oftype\sigma}{\Delta}{t}{\tau}$, $X$ a
t.c.\ denotation of type~$\tcden\tau$ 
w.r.t.\ $\tcden{\Gamma,v\oftype\sigma;\Delta}$, and
$t\apprby X$ then $\lambda v.t\apprby\llambda_\star v.X$.
\item \label{item:app}
If $\typing \Gamma{\Delta_0}s{\sigma\arrow\tau}$, 
$\typing \Gamma{\Delta_1}t\sigma$,
$\Delta_0$ and $\Delta_1$ satisfy the side-conditions of~\InfRule{$\arrow$-E},
$X$ and $Y$ are t.c.\ denotations of type $\tcden{\sigma\arrow\tau}$
and $\tcden\sigma$ w.r.t.\ $\tcden{\Gamma;\Delta_0}$ and
$\tcden{\Gamma;\Delta_1}$ respectively, and
$s\apprby X$ and $t\apprby Y$, then $st\apprby X\star Y$.
\end{enumerate}
\end{lem}
\begin{proof}
Part~\ref{item:easy} is a direct unwinding of the definitions and
Parts~\ref{item:abs} and~\ref{item:app} take a little more work.  The
details are essentially identical to those of the corresponding
induction steps of the proof of Lemma~70(b) in~\ATS.
\end{proof}

\begin{prop}
\label{safety-preservation}
If $X$, $Y$, and $Z$ are safe t.c.\ denotations of appropriate types,
then $\tcdcons(X)$, $\tcddest(X)$, $\tcdtest(X)$,
$\tcdcond(X,Y,Z)$, $\tcddown(X, Y)$, $\llambda_\star v.X$, and
$X\star Y$ are safe t.c.\ denotations.
\end{prop}
\begin{proof}[\proofsketch]
This is again an unwinding of definitions; we present the $X\star Y$ case
as an example.  
Suppose $X$ and $Y$ are t.c.\ denotations of type
$\tcden{\sigma\arrow\tau}$ and $\tcden{\sigma}$ respectively w.r.t.\ $\Sigma$,
$X\leq (P_X,p_X)$ and $Y\leq (P_Y,p_Y)$, where 
$\tail\tau = \base b$ and $p_X\oftype\potden{\sigma\arrow\tau} =
\potden\sigma\arrow\tcden\tau$ is 
$\base b$-safe.  By definition $\pot(p_Xv)$ is $\base b$-safe
w.r.t.~$\Sigma,v\oftype\potden{\sigma}$ where $v$ is a fresh variable.
By Lemma~32 of~\ATS\ (Substitution of
safe polynomials), $\pot(p_Xp_Y) \leq p$ for some $\base b$-safe
polynomial~$p$.
Since $\pot(X\star Y)\leq \pot(p_Xp_Y)$ we conclude that
$X\star Y$ is $\base b$-safe.\footnote{This and other similar computations
of the full proof rely on simple properties of the well-tempered 
semantics of~\ATS\ to which we alluded earlier.}
\end{proof}

\begin{thm}
\label{clm:atr-minus-soundness-prelim}
For every $\ATRm$ term~$\GDtyping t\tau$ there is a 
$\tail(\tcden\tau)$-safe t.c.\ denotation~$X$ of type~$\tcden\tau$
w.r.t.\ $\tcden{\Gamma;\Delta}$ such that $t\apprby X$.
\end{thm}
\begin{proof}
The proof is by induction on the typing inference.  The cases of the
induction step corresponding to the syntax-directed rules
are given by Lemma~\ref{atr-minus-soundness-istep} and
if the last line of the typing inference
is either \InfRuleP{Shift} or \InfRuleP{Subsumption}, then the
corresponding typing rule for t.c.\ polynomials applies.
So we are just left with establishing the base cases.  The constants
are easy and $x\apprby (x_c,x_p)$ by definition of $\rho\apprby\varrho$.
That leaves us with oracles.  We can give an explicit definition
of a safe t.c.\ denotation $\tcden\alpha$ such that 
$\alpha\apprby\tcden\alpha$ in terms of the \emph{length} of~$\alpha$.
However, defining the length of~$\alpha$
entails defining the \emph{length-types}, which would take
us somewhat far afield.  We delay these definitions until
Section~\ref{sec:polynomial-bounds},
when we show how to extract second-order polynomial bounds on the
cost of evaluating $\ATR$ programs.
\end{proof}

\begin{defn}
For an $\ATRm$ term $\GDtyping t\tau$ we define the
\emph{time-complexity interpretation of~$t$}, $\tcden t$, to be the
t.c.\ denotation~$X$ of Theorem~\ref{clm:atr-minus-soundness-prelim}.%
\footnote{Formally, of course, we should write $\tcden{\GDtyping t\tau}$,
but the typing should always be clear from context.}
\end{defn}

\begin{cor}[Soundness for $\ATRm$]
\label{clm:atr-minus-soundness}
For every $\ATRm$ term $\GDtyping t\tau$, $\tcden t$
is $\tail(\tcden\tau)$-safe w.r.t. $\tcden{\Gamma;\Delta}$
and $t\apprby\tcden t$.
\end{cor}

\section{Soundness for $\ATR$}
\label{sec:atr-soundness}

Our goal in this section is to extend the Soundness argument for~$\ATRm$
to handle $\crec$ terms, thereby proving Soundness for $\ATR$.
First we define \emph{plain affine recursion}
in Section~\ref{sec:plain-affine-recursion},
which captures (up to~$\eta$-equivalence)
how a recursively-defined function can occur in its definition.
In Section~\ref{sec:decomp-lemma}
we prove the Decomposition Lemma (Theorem~\ref{decomposition-lemma}), which
characterizes the t.c.\ denotations that bound plain affine
recursive definitions.
Specifically, we give an algebraic characterization in which the cost of the
application of the affine variable
occurs as a linear term with coefficient~$1$ (hence our terminology).
In Section~\ref{sec:solution-lemma} we use the Decomposition Lemma to
prove the Unfolding Lemma
(Theorem~\ref{solution-lemma} and Corollary~\ref{solution-lemma-as-poly}),
which gives polynomial bounds on recursively-defined functions in
terms of their recursion depth (Definition~\ref{defn:recn-depth}).
We also prove the Termination Lemma (Theorem~\ref{termination-lemma})
which gives polynomial bounds on the recursion depth.
This provides the last step needed to
prove Soundness for~$\ATR$ (Theorem~\ref{soundness-prelim}
and Corollary~\ref{soundness}).

\subsection{Plain affine recursion}
\label{sec:plain-affine-recursion}

As already noted, our list-operation and sorting programs use several
forms of recursion that go beyond tail recursion.  However, they
all boil down to (essentially) filling in the argument positions
of the recursively-defined function, then using the result in 
basic operations or as an argument to an application.  In
fact, they are all instances of the scheme of \emph{plain affine
recursion}:

\begin{defn}
Suppose that $\typing\Gamma{f\oftype\base b_1\arrow\dots\arrow\base b_k\arrow\base b_0}{t}{\base b}$.
$t$ is a \emph{plain affine recursive definition of~$f$}, or
\emph{$f$ is in plain affine position in~$t$}, if:
\begin{enumerate}
\item $f\notin\fv(t)$; or
\item $t = ft_1\dots t_k$ where $f\notin\fv(t_i)$ for any~$i$ (we call
this a \emph{complete application of~$f$}); or
\item $t = \cond{s}{s_0}{s_1}$ where $f\notin\fv(s)$ and 
each $s_i$ is a plain affine recursive definition of~$f$; or
\item $t=\comb{op} s$ where $\comb{op}$ is any of $\cons_a$, $\destr$,
or~$\test_a$ and $s$ is a plain affine recursive definition of~$f$; or
\item $t=\down s_0s_1$ where $s_0$ is a plain affine recursive definition
of~$f$ and
$f\notin\fv(s_1)$; or
\item $t = st_1\dots t_m$ where $f\notin\fv(s)$ and there is~$i$ such
that $t_i$ is a
plain affine recursive definition of~$f$ and $f\notin \fv(t_j)$ for $j\not=i$; 
or
\item $t = (\lambda x_1\dots x_m.s)t_1\dots t_m$ 
where $s$ is a plain affine recursive
definition of~$f$ and $f\notin\fv(t_i)$ for any~$i$ (we call this
a \emph{\keyw{let}-binding}).
\end{enumerate}
\end{defn}

Whereas in~\ATS\ we enforced a side condition on~\InfRuleP{\keyw{crec}-I}
that the recursively-defined function be in tail position, it would be
much nicer to be able to say that
if $\typing\Gamma{f\oftype\gamma}{t}{\base b}$, then $f$
occurs in plain affine position in~$t$.  As stated, this does not quite
hold.  An exception is
$(\lambda x.fs)t_1t_2$, which
is typeable with $f\oftype\base b_1\arrow\base b_2\arrow\base b$
from appropriate typings of~$s$, $t_1$, and $t_2$; but $f$ is not
in plain affine position in this expression.  
A trivial syntactic change ``fixes'' this expression
without changing the meaning: simply replace
$\lambda x.fs$ with $\lambda xy.fsy$ where $y$ is a fresh variable.  
In fact, it is not hard to show that 
this exception illustrates 
essentially the only way in which $f$ can occur affinely
in a term without being in plain affine position.

More precisely, we
define a recursive operation on base-type terms $t\mapsto t^\dagger$
as follows.
If $t=\cons_0s$ then $t^\dagger = \cons_0s^\dagger$, and the
operation ``pushes through'' $\cons_1$, $\destr$, $\test_b$, $\keyw{if}$, and
$\down$ similarly.
Assume we have a term~$t$ such that
$\typing\Gamma{f\oftype\gamma}t{\base b}$ where
$\gamma = \base b_1\arrow\dots\arrow\base b_k\arrow\base b_0$.  Consider
any base-type subterm of the form $ss_1\dots s_m$ that is not an immediate
subterm of an application and for which $s$ is not an application.  If
$f\in\fv(s_i)$ then necessarily $s_i$ is of base type, so 
$ss_1\dots s_{i-1}s_i^\dagger s_{i+1}\dots s_{m}$ is a plain affine definition
of~$f$.  If $f\in \fv(s)$, then $f\notin\fv(s_i)$ for any~$i$ and
$s$ cannot be a $\crec$-term, so $s$ has the form
$(\lambda x_1\dots x_i.s')$ for some~$i$ where $s'$ is not an
abstraction.  Replace $s$
with $(\lambda x_1\dots x_m.(sx_{i+1}\dots x_m))^\dagger$;
note that we have ``filled out'' the arguments of $s$ so that
$sx_{i+1}\dots x_m$ is of base type.
Of course, a formal definition
would impose an appropriate measure on terms and define~$t^\dagger$ recursively
in terms of that measure; we leave the details to the interested reader.
The relevant properties are as follows, all of which are easily verified
by unwinding the definitions:

\begin{prop}
\label{t-dagger-properties}
Suppose that $\typing\Gamma{f\oftype\gamma}{t}{\base b}$.  Then:
\begin{enumerate}
\item \label{type} $\typing\Gamma{f\oftype\gamma}{t^\dagger}{\base b}$.
\item \label{position} $f$ is in plain affine recursive position in $t^\dagger$.
\item \label{eval} For any environment~$\rho$, $\cl t\rho\evalto\cl z\theta$ iff
$\cl {t^\dagger}\rho\evalto\cl z\theta$.
\item \label{bound} If $t^\dagger\apprby X$ then $t\apprby X$.
\end{enumerate}
In particular, we can w.l.o.g.\ assume that the body of every
$\crec$ expression is a plain affine recursive definition.
\end{prop}

The next proposition shows that typing derivations of plain
affine recursive definitions can placed in a normal form.
We will use this normal form in our proof of the Decomposition
Lemma (Theorem~\ref{decomposition-lemma}), which characterizes the
t.c.\ denotations that bound plain affine recursive definitions.
We call the premis of~\InfRule{$\arrow$-E} that types the operator
the \emph{major} premis of the rule.

\begin{prop}
\label{par-normal-form}
Suppose $\mathcal D$ is a derivation of 
$\typing\Gamma{\Delta,f\oftype\gamma}t{\base b}$ where
$t$ is a plain affine definition of~$f$, $f\in\fv(t)$, and
$\gamma=\base (b_1,\dots,\base b_k)\arrow\base b_0$.  Then:
\begin{enumerate}
\item No \InfRuleP{Subsumption} inference is the last line of the major
premis of an \InfRuleP{$\arrow$-E} inference in which $f$ occurs free.
\item No \InfRuleP{Subsumption} inference immediately follows
an \InfRuleP{$\arrow$-I} inference in which $f$ occurs free.
\end{enumerate}
\end{prop}
\begin{proof}
The proof is by induction on the shape of~$t$ and we consider the possible
typings of each shape in turn.  The cases in which the induction hypothesis
does not immediately apply are $t=ft_1\dots t_k$ and
$t = (\lambda x_1\dots x_m.s)t_1\dots t_m$.

Suppose $t=ft_1\dots t_k$; for concreteness we take $k=2$ and we write
$\Sigma$ for $\Gamma;\Delta,f\oftype\gamma$.
Then $\mathcal D$ has the following general form:%
\footnote{It is here that we use the restriction that \InfRuleP{Shift} cannot
be applied if the affine zone is non-empty; without this restriction,
we could have a sequence of \InfRuleP{Shift} and \InfRuleP{Subsumption}
inferences interleaved with the \InfRuleP{$\arrow$-E} inferences, and
this proof would not carry through.}
\begin{prooftree}
\AXC{}
\UIC{$\tctyping\Sigma f\base b_1\arrow\base b_2\arrow\base b_0$}
\LeftLabel{\InfRule{Subsumption}}
\UIC{$\tctyping\Sigma f{\base b_1'\arrow\base b_2'\arrow\base b_0'}$}
\AXC{$\ityping\Gamma{t_1}{\base b_1'}$}
\BIC{$\tctyping\Sigma{ft_1}{\base b_2'\arrow\base b_0'}$}
\LeftLabel{\InfRule{Subsumption}}
\UIC{$\tctyping\Sigma{ft_1}{\base b_2''\arrow\base b_0''}$}
\AXC{$\ityping\Gamma{t_2}{\base b_2''}$}
\BIC{$\tctyping\Sigma{ft_1t_2}{\base b_0''}$}
\end{prooftree}
Since $\base b_1'\subtype\base b_1$, $\base b_2''\subtype\base b_2'\subtype\base b_2$, and $\base b_0\subtype\base b_0'\subtype\base b_0''$, we can
rewrite this derivation as
\begin{prooftree}
\AXC{}
\UIC{$\tctyping\Sigma f\gamma$}
\AXC{$\ityping\Gamma{t_1}{\base b_1'}$}
\LeftLabel{\InfRule{Subsumption}}
\UIC{$\ityping\Gamma{t_1}{\base b_1}$}
\BIC{$\tctyping\Sigma {ft_1} {\base b_2\arrow\base b_0}$}
\AXC{$\ityping\Gamma{t_2}{\base b_2''}$}
\LeftLabel{\InfRule{Subsumption}}
\UIC{$\ityping\Gamma{t_2}{\base b_2}$}
\BIC{$\tctyping\Sigma {ft_1t_2} {\base b_0}$}
\UIC{$\tctyping\Sigma {ft_1t_2} {\base b_0''}$}
\end{prooftree}

If $t=(\lambda x_1\dots x_m.s)\vec t$
then first apply the induction hypothesis to
the typing of~$s$.  
Any \InfRuleP{Subsumption} inferences that follow one of the
\InfRuleP{$\arrow$-I} inferences can be moved to the end of all
those inferences.
Thus as in the previous case, we can move any
\InfRuleP{Subsumption} inferences that occur as the last line of
a major premis in one of the~\InfRuleP{$\arrow$-E} inferences
$(\lambda\vec x.s)t_1\dots t_i$ to the minor premis, concluding with a possible
last \InfRuleP{Subsumption} inference.
\end{proof}

The \keyw{let}-binding clause of plain affine recursion leads us
to consider t.c.\ denotations of the form $(\llambda_\star\vec x.X)\star\vec Y$,
so we characterize them here.  First we define a function on
t.c.\ denotations that allows us to neatly express the ``overhead cost''
of combining t.c.\ denotations:

\begin{defn}
For any t.c.\ denotation~$X$, 
\[
  \dally(m,\;X) \eqdef \llambda\varrho\bigl(m+\tccost(X\varrho),\;\pot(X\varrho)\bigr).
\]
\end{defn}

\begin{prop}
\label{dally-preserves-safety}
If $X$ is a safe t.c.\ denotation, then so is $\dally(m, X)$.
\end{prop}

\begin{prop}
\label{abs-app-tc}
Let $X$ be a t.c.\ denotation 
w.r.t.\ $\Sigma,\tcden{x_1\oftype\sigma_1,\dots,x_m\oftype\sigma_m}$ 
and $Y_1,\dots,Y_m$ be t.c.~denotations w.r.t.\ $\Sigma$.  Then
\[
  (\llambda_\star\vec x.X)\star\vec Y = \llambda\varrho.\dally\bigl(2m+\sum_{i=1}^m\tccost(Y_i\varrho),\;X\extend\varrho{x_i}{\val(\pot(Y_i\varrho))}\bigr).
\]
\end{prop}
\begin{proof}
The proof is by induction on~$m$; the base case is immediate.
For the induction step we apply the induction hypothesis and
unwind definitions.  In the following calculation
we write $Y_{ic}\varrho$ for $\tccost(Y_i\varrho)$,
$\varrho_m$ for
$\extend\varrho{x_i}{\val(\pot(Y_i\varrho))}$ where $i=1,\dots,m$,
and similarly for $\varrho_{m+1}$:
\begin{align*}
(\llambda_\star &x_1\dots x_{m+1}.X)\star Y_1\star\dots\star Y_{m+1} \\
  &= \bigl(\llambda\varrho.\dally\bigl(2m+\sum_{i=1}^m Y_{ic}\varrho,\;(\llambda_\star x_{m+1}.X)\varrho_m\bigr)\bigr)\star Y_{m+1} \\
  &= \Bigl(\llambda\varrho.\dally\Bigl(2m+\sum_{i=1}^m Y_{ic}\varrho, \\
  &\qquad\bigl(\llambda\varrho'\bigl(1,\;\llambda x_{m+1,p}.X\extend{\varrho'}{x_{m+1}}{\val(x_{m+1,p})}\bigr)\bigr)\varrho_m\Bigr)\Bigr)\star Y_{m+1} \\
  &= \llambda\varrho.\bigl(1+2m+\sum_{i=1}^m Y_{ic}\varrho + 1+ Y_{m+1,c}\varrho + \tccost(X\varrho_{m+1}),\; \pot(X\varrho_{m+1})\bigr) \\
  &= \llambda\varrho.\dally\bigl(2(m+1)+\sum_{i=1}^{m+1}Y_{ic}\varrho,\;
     X\varrho_{m+1}\bigr).
\end{align*}
\end{proof}

\subsection{Bounds for recursive definitions: the Decomposition Lemma}
\label{sec:decomp-lemma}

We now state and prove the Decomposition Lemma.  
Throughout this section and
the next we will need to assume that induction hypothesis of the Soundness 
Theorem holds, 
because the Decomposition Lemma will be used in its induction step.  
So to shorten the statements of the coming claims, we name the
induction hypothesis:
\begin{description}
\item[Inductive Soundness Assumption (ISA)]
A term $\typing\Gamma{f\oftype\gamma}t{\base b}$
(where $\gamma = (\base b_1,\dots,\base b_k)\arrow\base b_0$)
satisfies the \emph{inductive Soundness assumption} if $t$ is a plain
affine recursive definition of~$f$
and whenever $\ityping{\Gamma'}s\tau$ is a subterm of~$t$, there
is a $\tail(\tcden\tau)$-safe 
t.c.\ polynomial $(P_s,p_s)$ w.r.t.~$\tcden{\Gamma'}$ such that
$s\bddby(P_s,p_s)$.
\end{description}

For the statement of the Decomposition Lemma,
recall our convention that in writing a t.c.\ polynomial~$p$
w.r.t.~$\tcden{\Gamma;\Delta}$, if $x\in\dom(\Gamma\union\Delta)$ we
write $p(\dots,x,\dots)$ to abbreviate $p(\dots,x_c,x_p,\dots)$.

\begin{thm}[Decomposition Lemma]
\label{decomposition-lemma}
Suppose $\typing\Gamma{f\oftype\gamma}t{\base b}$ satisfies the ISA
and that $\dom\Gamma=\vec y$.  Then
\[
  t\apprby \bigl(P(\vec y,\pot(f\star\vec p))+\tccost(f\star\vec p),\;p(\vec y,\pot(f\star\vec p))\bigr)
\]
where $P(\vec y,w^{\potden{\base b_0}})\oftype\Tally$ is a cost polynomial,
$p(\vec y,w^{\potden{\base b_0}})\oftype\potden{\base b}$ is a
$\potden{\base b}$-safe potential polynomial, and
$\vec p = p_1,\dots,p_k$ where for each~$i$,
$p_i = p_i(\vec y)\oftype\tcden{\base b_i}$ is a
$\tail(\tcden{\base b_i})$-safe t.c.\ polynomial.\footnote{Recall from
Proposition~\ref{no-cost-vars-in-pot-polys}
that since $p$ is a potential
polynomial, we can in fact assume that $p(\vec y,w) = p(\dots,y_{ip},\dots,w)$.}
If $f\notin\fv(t)$, read $f\star\vec p$ as $(0,0)$.
\end{thm}
\begin{proof}
The proof is by induction on the typing of~$t$.  For clarity
we drop mention of the parameters~$\vec y$ everywhere.  If $f\notin\fv(t)$, then
the claim follows from the ISA.  Also notice
that if the last line of the typing of~$t$ is \InfRuleP{Subsumption} then
the claim follows immediately from the induction hypothesis, because
if $\base b'\subtype\base b$, then any $\potden{\base b'}$-safe polynomial
is $\potden{\base b}$-safe.  The last line cannot be
\InfRuleP{Shift} because this rule cannot be applied to a
judgment with non-empty affine zone.

If the last line of the typing is \InfRuleP{\keyw{op}-I},
\InfRuleP{\keyw{if}-I}, or \InfRuleP{$\down$-I}
then the claim follows from the induction
hypothesis by using the appropriate operation from 
Figure~\ref{fig:atr-minus-tc-ops}; we present the \InfRuleP{\keyw{if}-I} case
as an example.
Suppose the last line of the typing is $\InfRuleP{\keyw{if}-I}$, so that
$t = \cond{s}{t_0}{t_1}$.  By the ISA we have
that $s\bddby(P_s,p_s)$, and by the induction hypothesis that
$t_i\bddby(P^i(\pot(f\star\vec{p^i}))+\tccost(f\star\vec{p^i}),\,
p^i(\pot(f\star\vec{p^i})))$ for appropriate polynomials $P^i$, $p^i$,
and $\vec{p^i}=p^i_1,\dots,p^i_k$.  
By Lemma~\ref{atr-minus-soundness-istep} we have that 
\begin{align*}
t & \bddby \Bigl(1+P_s+
  \bigr(\bigl(P^1(\pot(f\star\vec{p^1}))+\tccost(f\star\vec{p^1})\bigr)\bmax \bigl(P^2(\pot(f\star\vec{p^2}))+\tccost(f\star\vec{p^2})\bigr)\bigr), \\
  & \phantom{\bddby(}\qquad p^1(\pot(f\star\vec{p^1}))\bmax p^2(\pot(f\star\vec{p^2}))\Bigr) \\
  & \leq \bigl(1+P_s+P(\pot(f\star\vec p))+\tccost(f\star\vec{p}),\;p(\pot(f\star\vec p))\bigr)
\end{align*}
where $P = P^1\bmax P^2$, $p$ is a safe t.c.\ polynomial greater than
$p^1\bmax p^2$, and $p_i$ is a safe t.c.\ polynomial greater 
than $p^0_i\bmax p^1_i$ (see Proposition~\ref{bound-max-of-poly}).

The only other possibility is that the last line
is \InfRuleP{$\arrow$-E}, and for that we break into cases depending
on the exact form of~$t$.

\textsc{Case 1:}\enspace $t=ft_1\dots t_k$.
By Proposition~\ref{par-normal-form}
we can assume that we have typings $\ityping\Gamma{t_i}{\base b_i}$.
Since $f\notin\fv(t_i)$ we have $\tcden{\base b_i}$-safe t.c.\ polynomials
$p_i$ such that $t_i\apprby p_i$ and it follows from
Lemma~\ref{atr-minus-soundness-istep} that
$t\apprby f\star\vec p = (\tccost(f\star\vec p),\, \pot(f\star\vec p))$.

\textsc{Case 2:}\enspace
$t=st_1\dots t_m$ where w.l.o.g.\ $t_m$ is a plain
affine definition of~$f$ and $f\in\fv(t_m)$.
We can assume that 
$\ityping\Gamma{st_1\dots t_{m-1}}{\base b'\arrow\base b}$ and
$\typing\Gamma{f\oftype\gamma}{t_m}{\base b'}$ for some $\base b'$.
Since $f\notin\fv(st_1\dots t_{m-1})$ the ISA tells us that
$st_1\dots t_{m-1} \apprby (P_s, p_s)\oftype\tcden{\base b'\arrow\base b}$
where $(P_s,p_s)$ is $\potden{\base b}$-safe.  The induction hypothesis
tells us that
$t_m\apprby(P(\pot(f\star\vec p))+\tccost(f\star\vec p),\,p(\pot(f\star\vec p)))$
so by Lemma~\ref{atr-minus-soundness-istep} we conclude that
\[
  t\apprby \bigl(1+P_s+P(\pot(f\star\vec p))+\tccost(f\star\vec p)+\tccost(p_s(p(\pot(f\star\vec p)))), \\ \pot(p_s(p(\pot(f\star\vec p))))\bigr).
\]
Since $p_s\oftype\potden{\base b'}\arrow\tcden{\base b}$ is 
$\potden{\base b}$-safe and $p(w^{\potden{\base b_0}})\oftype\potden{\base b'}$
is $\potden{\base b'}$-safe, we have that
$p_s(p(\pot(f\star\vec p)))\oftype\tcden{\base b}$ is
$\potden{\base b}$-safe,\footnote{Actually, bounded by a $\potden{\base b}$-safe
polynomial; from now on we shall assume that the reader can insert the
``bounded by'' qualification as needed.} and hence that
$\pot(p_s(p(\pot(f\star\vec p))))$ is $\potden{\base b}$-safe, completing
the proof for this case.

\textsc{Case 3:}\enspace $t=(\lambda x_1\dots x_m.s)t_1\dots t_m$
where $s$ is a plain affine definition of~$f$.  By
Proposition~\ref{par-normal-form} we may assume that we have typings
$\typing{\Gamma,\vec x\oftype\vec\sigma}{f\oftype\gamma}{s}{\base b}$
and $\ityping\Gamma{t_i}{\sigma_i}$.  The induction hypothesis
tells us that
$s\apprby(P_s(\vec x,\pot(f\star\vec p))+\tccost(f\star\vec p),\,p_s(\vec x,\pot(f\star\vec p)))$ where
$p_i = p_i(\vec x)$ and
the ISA tells us that $t_i\apprby(P^i,p^i)$.
Using Lemma~\ref{atr-minus-soundness-istep} and Proposition~\ref{abs-app-tc}
we conclude that
\begin{multline*}
t\apprby \Bigl(2m+\sum_{i=1}^m P^i + P_s\bigl(\val(p^1),\dots,\val(p^m),\pot(f\star\vec{p'})\bigr)+\tccost(f\star\vec{p'}),\\
p_s\bigl(p^1,\dots,p^m,\pot(f\star\vec{p'})\bigr)\Bigr)
\end{multline*}
where $p_i' = p_i(\val(p^1),\dots,\val(p^m))$.
Since each $p^i\oftype\potden{\sigma_i}$ is $\tail(\potden{\sigma_i})$-safe,
$p_i'$ is $\potden{\base b_i}$-safe, and substuting safe polynomials into
safe polynomials yields a t.c.\ denotation that is bounded by a safe
polynomial (\ATS\ Lemma~32), the claim is established.
\end{proof}

\subsection{Polynomial bounds for recursive terms}

\subsubsection{Bounds in terms of recursion depth:  the Unfolding Lemma.}
\label{sec:solution-lemma}

From the Decomposition Lemma we know that if
$\typing{\Gamma,\vec v\oftype\vec{\base b}}{f\oftype\gamma} t{\base b}$ 
satisfies the ISA, then
\[
  t \bddby \bigl(P(\vec v,\pot(f\star\vec p))+\tccost(f\star\vec p),\;q\pmjb(r\bmax\pot(f\star\vec p))\bigr)
\]
where $q = q(\vec v)$ is $\potden{\base b}$-strict and
$r = r(\vec v)$ is $\potden{\base b}$-chary (we have supressed mention of
the variables other than $\vec v$ and $f$).  Let $X_t$ denote this
t.c.\ denotation.  
Also define the (syntactic) substitution
function
\[
  \xi_t= [\tccost(\val(p_{ip})), \pot(\val(p_{ip}))/v_{ic},v_{ip}]
\]
and set
$\xi_t^0 = \id$ and $\xi_t^{n+1} = \xi_t^n\comp\xi_t$ (we write the syntactic
substitution of the polynomial $p$ for the variable~$x$ in the
t.c.\ denotation~$X$ by $\subst X x p$).  The point behind
these functions is that if $p(v_1,\dots,v_k)$ is a polynomial, then
\[
  (\llambda_\star v_1\dots v_k.p)\star p_1\star\dots\star p_k = \dally\Bigl(2k+\sum_{i=1}^kp_{ic},\; p\xi_t\Bigr)
\]
by Proposition~\ref{abs-app-tc} and expressions of this form
arise frequently in our analysis.

To analyze the of closures of the form
$\cl t{\extend\rho{f}{\cl*{\crec(\bz^\ell)(\afflambda f.\lambda\vec v.t)}\rho}}$
where $t$ is a plain affine recursive definition of~$f$,
we will actually need to analyze subterms of~$t$ under
extensions of the environment indicated here.
To that end, we make some definitions in order to
simplify the statements of the coming claims.

\begin{defn}
Suppose $\typing{\Gamma,v_1\oftype\base b_1,\dots,v_k\oftype\base b_k}{f\oftype\gamma}t{\base b}$ satisfies
the ISA.  Define
\begin{enumerate}
\item $\Gamma_{\vec v} = \Gamma,v_1\oftype\base b_1,\dots,v_k\oftype\base b_k$;
\item $C_{t,\ell} \eqdef \crec(\bz^\ell)(\afflambda f.\lambda\vec v.t)$;
\item $T_{t,\ell} \eqdef \lambda\vec v.\cond{\lh{\bz^\ell}<\lh{v_1}}{t}{\eps}$;
\item For $\rho\in\Env{\Gamma_{\vec v}}$,
$\rho_{t,\ell} \eqdef \extend\rho{f}{\cl{C_{t,\ell}}{\rho}}$.  
\end{enumerate}
Notice that
$\cl{C_{t,\ell}}{\rho}\evalto\cl{T_{t,\ell}}{\rho_{t,\ell+1}}$ is an axiom
of the evaluation relation.
We write $\cl t{\rho_\ell}$ for $\cl{t}{\rho_{t,\ell}}$.
\end{defn}

\begin{defn}
\label{defn:recn-depth}
Suppose $\typing {\Gamma_{\vec v}}{f\oftype\gamma}t{\base b}$ 
satisfies the ISA,
$\typing{\Gamma^*}{f\oftype\gamma}{t^*}{\base b^*}$ is a subterm of~$t$,
$\rho\in\Env{\Gamma_{\vec v}}$, and
$\rho^*\in\Env{\Gamma^*}$ is an extension of~$\rho$.  
The \emph{recursion-depth} of~$\cl {t^*}{\rho^*_{t,\ell}}$,
$\rdp(\cl {t^*}{\rho^*_{t,\ell}})$ is defined to be the number of
$\crec$ axioms $\cl {C_{t,m}}{\rho}\evalto \cl{T_{t,m}}{\rho_{t,m+1}}$ in
the evaluation derivation of~$\cl{t^*}{\rho_{t,\ell}^*}$ when
$\cl{t^*}{\rho^*_{t,\ell}}\evalto\cl z\theta$ for some~$\cl z\theta$,
and $\rdp(\cl{t^*}{\rho^*_{t,\ell}}) = \infty$ otherwise.
\end{defn}

The Unfolding Lemma establishes bounds on evaluating closures in terms of
recursion depth.  The proof is a nested induction:  first on the recursion
depth, and then on the shape of the plain affine definition.  Because of the
many cases its length may hide the simplicity of what is going on, so we
make that explicit here:  a
careful calculation of the cost of one recursive call in the evaluation.

\begin{thm}[Unfolding Lemma]
\label{solution-lemma}
Suppose $\typing{\Gamma_{\vec v}}{f\oftype\gamma} t{\base b}$
satisfies the ISA.  Let  $\xi=\xi_t$ be given as above.
Suppose $\rho\in\Env{\Gamma_{\vec v}}$, 
$\varrho\in\Env{\tcden{\Gamma_{\vec v}}}$, $\rho\bddby\varrho$, and
that $\rdp(\cl t{\rho_\ell}) = d<\infty$.  Then:
\begin{enumerate}
\item If $\base b$ is computational,
\[
  \cl t{\rho_\ell} \bddby \bigl(d(10+3p_{1p})+(d+1)(2k+\sum_{i=1}^k p_{ic} + P(dq+r)),\;(d+1)q+r\bigr)\xi^d\varrho.
\]
\item If $\base b$ is oracular,
\[
  \cl t{\rho_\ell} \bddby \bigl(d(10+3p_{1p})+(d+1)(2k+\sum_{i=1}^k p_{ic} + P(q\bmax r)),\;q\bmax r\bigr)\xi^d\varrho.
\]
\end{enumerate}
\end{thm}
\begin{proof}
The proof is by induction on~$d$.  For the base case ($d=0$) we
prove the following claim:
\begin{quotation}
Suppose $\typing{\Gamma^*_{\vec v}}{f\oftype\gamma}{t^*}{\base b^*}$ is
a subterm of~$t$ and take
$X^*$ so that $t^*\bddby X^*$ by the Decomposition Lemma.  Suppose
$\rho^*\in\Env{\Gamma^*_{\vec v}}$ is an extension of~$\rho$,
$\varrho^*\in\Env{\tcden{\Gamma^*_{\vec v}}}$ is an extension of~$\varrho$, and
$\rho^*\bddby\varrho^*$.
If $\rdp(\cl {t^*}{\rho^*_{t,\ell}})=0$ then
$\cl{t^*}{\rho^*_{t,\ell}}\bddby \subst{X^*}{f}{\tcz}\varrho^*$
where $\tcz = \llambda_\star\vec v.(0,0)$.
\end{quotation}
First let us see that this claim yields the desired bound when $d=0$.
It tells us that 
$\cl{t}{\rho_\ell}\bddby\subst{X_t}{f}{\tcz}\varrho$.  Thus if $\base b$
is computational
\begin{align*}
\cl t{\rho_\ell}
  & \bddby \subst{\bigl(P(\pot(f\star\vec p))+\tccost(f\star\vec p),\;q+(r\bmax\pot(f\star\vec p))\bigr)}{f}{\tcz}\varrho \\
  &= \bigl(P(0)+(2k+\sum p_{ic}),\; q+(r\bmax 0)\bigr)\varrho \\
  & \leq \bigl(2k+\sum p_{ic} + P(r),\; q+r\bigr)\xi^0\varrho.
\end{align*}
The calculation is similar when $\base b$ is oracular.

We prove the claim by induction on the shape of~$t^*$ (a plain affine
definition of~$f$ that satisfies the ISA).  
For each case of the induction, we import the
notation from the corresponding case in the proof of the Decomposition
Lemma.  
We give the details for
a few cases, leaving the rest to the reader.
The case in which $t^* = ft_1\dots t_k$ is not possible, because
necessarily $\rdp(\cl*{ft_1\dots t_k}{\rho^*_{t,\ell}}) > 0$.  

\textsc{Case 1:}\enspace
$t^* = \cond{s}{t_0}{t_1}$.  Consider the subcase in which
$\cl{s}{\rho^*_{t,\ell}}\evalto \eps\theta$ (the other subcase is analogous).  
An analysis of the evaluation
of $\cl{t^*}{\rho^*_{t,\ell}}$ yields
\begin{align*}
\cost(\cl{t^*}{\rho^*_{t,\ell}})
  &= 1 + \cost(\cl s{\rho^*_{t,\ell}}) + \cost(\cl{t_0}{\rho^*_{t,\ell}}) \\
  &\leq 1 + P_s\varrho^* + \tccost(\subst{X_{t_0}}{f}{\tcz}\varrho^*) \\
\intertext{(by applying the ISA to $s$ and
secondary induction hypothesis to $t_0$)}
  &\leq (1+P_s + (\tccost(\subst{X_{t_0}}{f}{\tcz})\bmax\tccost(\subst{X_{t_1}}{f}{\tcz})))\varrho^* \\
  &= \tccost(\subst{X^*}{f}{\tcz}\varrho^*).
\end{align*}
Furthermore, if $\cl{t^*}{\rho^*_{t,\ell}}\evalto\cl z\theta$ then
$\cl{t_0}{\rho^*_{t,\ell}}\evalto\cl z\theta$, so again by the secondary
induction hypothesis we have that
\[
\cl z\theta
  \bddbypot \pot(\subst{X_{t_0}}{f}{\tcz})\varrho^* 
  \leq \pot(\subst{X_{t_0}}{f}{\tcz}\bmax \subst{X_{t_1}}{f}{\tcz})\varrho^* 
  = \pot(\subst{X^*}{f}{\tcz}\varrho^*).
\]
The two facts together tell us that
$\cl{t^*}{\rho^*_{t,\ell}}\bddby\subst{X^*}{f}{\tcz}\varrho^*$.

\textsc{Case 2:}\enspace
$t^* = st_1\dots t_m$ where w.l.o.g.\ $t_m$ is a plain affine definition
of~$f$ and $f\in\fv(t_m)$.  By the secondary induction
hypothesis we may assume that
$\cl{t_m}{\rho^*_{t,\ell}}\bddby \subst{X_{t_m}}{f}{\tcz}\varrho^*$ 
and following the
notation of the Decomposition Lemma
$st_1\dots t_{m-1}\bddby(P_s,p_s)$.
Suppose $\cl*{st_1\dots t_{m-1}}{\rho^*_{t,\ell}}\evalto\cl*{\lambda x.s'}{\theta'}$
(the case of evaluating to an oracle is similar),
$\cl{t_m}{\rho^*_{t,\ell}}\evalto\cl{z''}{\theta''}$, and
$\cl{s'}{\extend{\theta'}{x}{\cl{z''}{\theta''}}}\evalto\cl z\theta$ 
(these evaluations
are all defined because they are subevaluations of that of
$\cl{t^*}{\rho^*_{t,\ell}}$).
By definition of~$\bddby$ we have that
$\cl{s'}{\extend{\theta'}{x}{\cl{z''}{\theta''}}}\bddby 
p_s(\pot(\subst{X_{t_m}}{f}{\tcz}\varrho^*))$.
An analysis of the evaluation of $\cl{t^*}{\rho^*_{t,\ell}}$ yields
\begin{align*}
\cost(\cl{t^*}{\rho^*_{t,\ell}})
  & = 1 + \cost(\cl*{st_1\dots t_{m-1}}{\rho^*_{t,\ell}}) + \cost(\cl{t_m}{\rho^*_{t,\ell}}) + \cost(\cl {s'}{\extend{\theta'}{x}{\cl{z''}{\theta''}}}) \\
  &\leq 1 + P_s + \tccost(\subst{X_{t_m}}{f}{\tcz}) + \tccost(p_s(\pot(\subst{X_{t_m}}{f}{\tcz}\varrho^*))) \\
  &= \tccost(\subst{X^*}{f}{\tcz}\varrho^*).
\end{align*}
And if $\cl{t^*}{\rho^*_{t,\ell}}\evalto\cl{z}{\theta}$ then
$\cl{s'}{\extend{\theta'}{x}{\cl{z''}{\theta''}}}\evalto\cl z\theta$ so
we conclude that
\[
  \cl z\theta\bddbypot \pot(p_s(\pot(\subst{X_{t_m}}{f}{\tcz}\varrho^*))) =
  \pot(\subst{X^*}{f}{\tcz}\varrho^*).
\]

\textsc{Case 3:}\enspace
$t^* = (\lambda x_1\dots x_m.s)\vec t$ where $s$ is a plain affine definition of~$f$.
Say that
$\cl{t_i}{\rho^*_{t,\ell}}\evalto\cl{z_i'}{\theta_i'}$ and
$\cl{s}{\extend{\rho^*_{t,\ell}}{x_i}{\cl{z_i'}{\theta_i'}}}\evalto\cl z\theta$
(the evaluations of the subterms and body are all defined because
they are sub-evaluations of $\cl{t^*}{\rho^*_{t,\ell}}$).
Following the notation of the Decomposition Lemma we have
$t_i\bddby(P^i,p^i)$, so $\cl{z_i'}{\theta_i'}\bddbypot p^i$.
By the secondary induction hypothesis we have that
$\cl s{\extend{\rho^*_{t,\ell}}{x_i}{\cl{z_i'}{\theta_i'}}}\bddby
\subst{X_s}{f}{\tcz}\extend{\varrho^*}{x_i}{\val(p^i\varrho^*)}$.
An analysis
of the evaluation derivation of $\cl{t^*}{\rho^*_{t,\ell}}$ yields
\begin{align*}
\cost(\cl{t^*}{\rho^*_{t,\ell}})
  &= 2m+\sum\cost(\cl{t_i}{\rho^*_{t,\ell}}) + \cost(\cl s{\extend{\rho^*_{t,\ell}}{x_i}{\cl{z_i'}{\theta_i'}}}) \\
  &\leq 2m + \sum P^i\varrho^* + \tccost(\subst{X_s}{f}{\tcz}\extend{\varrho^*}{x_i}{\val(p^i\varrho^*)}) \\
\intertext{($\cost(\cl{t_i}{\rho^*_{t,\ell}}) = \cost(\cl{t_i}{\rho^*})$
because $f\notin\fv(t_i)$)}
  &= \subst{\bigl(2m + \sum P^i + P_s(\dots,\val(p^i),\dots,\pot(f\star\vec{p'}))+ \\
  &\qquad\qquad \tccost(f\star\vec{p'})\bigr)}{f}{\tcz}\varrho^* \\
  &= \tccost(\subst{X^*}{f}{\tcz}\varrho^*)
\end{align*}
where $p'_j = p_j(\dots,\val(p^i),\dots)$ and
\begin{align*}
\cl z\theta
  &\bddbypot \pot(\subst{X_s}{f}{\tcz}\extend{\varrho^*}{x_i}{\val(p^i\varrho^*)}) \\
  &= \subst{p_s(\vec x,\pot(f\star\vec{p}))}{f}{\tcz}\extend{\varrho^*}{x_i}{\val(p^i\varrho^*)} \\
  &= \subst{p_s(\dots,p^i,\dots,\pot(f\star\vec{p'}))}{f}{\tcz}{\varrho^*} \\
  &= \pot(\subst{X^*}{f}{\tcz}\varrho^*).
\end{align*}
Thus $\cl{t^*}{\rho^*_{t,\ell}}\bddby\subst{X^*}{f}{\tcz}\varrho^*$.
This completes the proof of the Unfolding Lemma.

For the induction step, suppose that $\rdp(\cl t{\rho_{t,\ell}}) = d+1$.
We show just the case when $\base b$ is computational; the oracular case
is similar.  Set
\[
  Y = \bigl(d(10+3p_{1p})+(d+1)(2k+\sum {p_{ic}} + P(dq+r)),\;(d+1)q+r\bigr)\xi^d.
\]
We will prove the following claim:
\begin{quotation}
Suppose $t^*$, $X^*$, $\rho^*$, and $\varrho^*$ are as in the claim
for the base case $d=0$ and suppose
$X^* = (P^*(\pot(f\star{\vec{p^*}}))+\tccost(f\star\vec{p^*}),\, p^*(\pot(f\star\vec{p^*})))$.  
If $\rdp(\cl{t^*}{\rho^*_{t,\ell}}) = d+1$
then $\cl{t^*}{\rho^*_{t,\ell}}\bddby\dally(10+3p_{1p}^*,\,\subst{X^*}{f}{\llambda_\star\vec v.Y})\varrho^*$.
\end{quotation}
Again we first show that this claim is sufficient for establishing
desired bound for the induction step.  From it we calculate
\begin{align*}
\cl{t}{\rho_{t,\ell}}
  &\bddby\dally\bigl(10+3p_{1p},\;\subst{X_t}{f}{\llambda_\star\vec v.Y}\bigr)\varrho \\
  &= \subst{\bigl(10+3p_{1p}+P(\pot(f\star\vec p))+\tccost(f\star\vec p), 
     q+(r\bmax\pot(f\star\vec p))\bigr)}{f}{\llambda_\star\vec v.Y}\varrho \\
  &\leq \bigl(10+3p_{1p}+P(\pot(Y\xi))+(2k+\sum p_{ic} + \tccost(Y\xi)),
        q+(r\bmax\pot(Y\xi))\bigr)\varrho \\
  &\leq \Bigl(10+3p_{1p}+P(((d+1)q+r)\xi^{d+1}) + 2k+\sum p_{ic} + \\
  &\phantom{\leq\Bigl(} \bigl(d(10 + 3p_{1p}) + (d+1)\bigl(2k+\sum p_{ic} + P(dq+r)\bigr)\bigr)\xi^{d+1}, \\
  &\phantom{\leq\Bigl(}\qquad q+(r\bmax((d+1)q+r)\xi^{d+1})\Bigr)\varrho \\
  &\leq \bigl((d+1)(10+3p_{1p}) + (d+2)(2k+\sum p_{ic} + P((d+1)q+r))
        (d+2)q+r\bigr)\xi^{d+1}\varrho
\end{align*}
using the fact that everything in sight is monotone and non-decreasing.

Establishing the claim is very similar to the $d=0$ claim; we present just one
key case here.  Suppose that $t^* = ft_1\dots t_k$ and
$\cl{t_i}{\rho^*_{t,\ell}}\evalto\cl{z_i}{\theta_i}$.  Also take
$p_i^*$ so that $t_i\bddby p_i^*$ so that $X^* = f\star\vec{p^*}$.  Then
analysing the evaluation derivation we have that
\[
  \cost(\cl{t^*}{\rho^*_{t,\ell}}) = (2+k+\sum p_{ic}^*\varrho^* + (8+3p_{1p}^*)\varrho^* + \cost(\cl{t}{\extend{\rho_{t,\ell+1}}{v_i}{\cl{z_i}{\theta_i}}})
\]
(the $8+3p_{1p}$ term is from the clock test when evaluating
$\cl{T_{t,\ell}}{\extend{\rho_{t,\ell+1}}{v_i}{\cl{z_i}{\theta_i}}}$).  Since
$\rdp(\cl{t^*}{\rho^*_{t,\ell}}) = d+1$ and the evaluation of
$\cl{t}{\extend{\rho_{t,\ell+1}}{v_i}{\cl{z_i}{\theta_i}}}$ is a subevaluation
we have 
that $\rdp(\cl{t}{\extend{\rho_{t,\ell+1}}{v_i}{\cl{z_i}{\theta_i}}}) = d$
and so the main induction hypothesis applies to let us conclude that
$\cl{t}{\extend{\rho_{t,\ell+1}}{v_i}{\cl{z_i}{\theta_i}}}\bddby
Y\extend{\varrho}{v_i}{\val(p^*_{ip}\varrho^*)}$.
Thus
\begin{align*}
\cost(\cl{t^*}{\rho^*_{t,\ell}})
  &\leq (2+k+\sum p_{ic}^* + (8+3p_{1p}^*) + \tccost(Y\extend\varrho{v_i}{\val(p_{ip}^*\varrho^*}))\varrho^* \\
  &= (10 + 3p_{1p}^* + \tccost(((\llambda_\star\vec v.Y)\star\vec {p^*})\varrho^*) \\
  &= \subst{(10+3p_{1p}^*+\tccost(f\star\vec {p^*}))}{f}{\llambda_\star\vec v.Y}\varrho^*.
\end{align*}
Furthermore, if $\cl{t^*}{\rho^*_{t,\ell}}\evalto\cl z\theta$ then
$\cl t{\extend{\rho_{t,\ell+1}}{v_i}{\cl{z_i}{\theta_i}}}\evalto\cl z\theta$
and so
\[
\cl z\theta \bddbypot \pot(Y\extend{\varrho}{v_i}{\val(p_{ip}^*\varrho^*}) = \pot((\llambda_\star\vec v.Y)\star\vec {p^*})\varrho^* = \\
\pot(\subst{(f\star\vec {p^*})}{f}{\llambda_\star\vec v.Y})\varrho^*.
\]
We conclude that
$\cl{t^*}{\rho^*_{t,\ell}}\bddby\dally(10+3p_{1p}^*,\subst{X^*}{f}{\llambda_\star\vec v.Y})\varrho^*$.
\end{proof}

\begin{cor}[Polynomial Unfolding Lemma]
\label{solution-lemma-as-poly}
Suppose $\typing{\Gamma_{\vec v}}{f\oftype\gamma}t{\base b}$ satisfies
the ISA, $\rho\in\Env{\Gamma_{\vec v}}$,
$\varrho\in\Env{\tcden{\Gamma_{\vec v}}}$, $\rho\bddby\varrho$.
Then there is a $\potden{\base b}$-safe time-complexity polynomial
$\phi(\vec v,d^{\potden{\base b_1}})$ such that for all~$\ell$ such that
$\rdp(\cl t{\rho_\ell})<\infty$, 
$\cl t{\rho_\ell}\bddby\phi(\vec v,\rdp(\cl t{\rho_\ell}))$.
\end{cor}
\begin{proof}
Using the Unfolding Lemma, it suffices to show that the map
$v_{ip}\xi_t^d$ is a safe polynomial 
w.r.t.\ $v_{ip}\oftype\potden{\base b_i}, 
d\oftype\potden{\base b_1}$.  This is precisely the content of the
One-step and $n$-step Lemmas of \ATS\ (Lemmas~$44$ and~$45$).
\end{proof}

\subsubsection{Bounds on recursion depth:  the Termination Lemma.}

Next we prove the Termination Lemma, which establishes a polynomial
bound on~$\rdp(\cl t{\rho_\ell})$; this will allow us to apply
the Unfolding Lemma.  Since we cannot \emph{a priori} assume that we
have an evaluation of $\cl t{\rho_\ell}$, we need a formalism that
allows us to refer to ``non-terminating evaluations.''  We sketch the idea here.
Introduce a new value $\unknown$.
Define the \emph{truncated evaluation relation}
$\cl s\rho\trevalto\cl z\theta$ just like the usual evaluation
relation~$\evalto$, but with an additional axiom:
\begin{prooftree}
\AXC{}
\UIC{$\cl*{\crec(\bz^\ell)(\afflambda f.\lambda\vec v.t)}{\rho}\trevalto\unknown$}
\end{prooftree}
Furthermore, for each inference rule of~$\evalto$ we add additional rules
that say that if one of the hypotheses evaluates to~$\unknown$, then
the remaining hypotheses (to the right) are ignored and the conclusion
evaluates to~$\unknown$.  For example, we have the additional inferences
\[
\AXC{$\cl r\rho\trevalto\unknown$}
\UIC{$\cl*{rs}{\rho}\trevalto\unknown$}
\DisplayProof
\qquad
\AXC{$\cl r\rho\trevalto\cl*{\lambda x.r'}{\theta'}$}
\AXC{$\cl s\rho\trevalto\unknown$}
\BIC{$\cl*{rs}{\rho}\trevalto\unknown$}
\DisplayProof
\]

We will use these truncated evaluations to establish a bound on the
recursion depth of ordinary evaluations.  The idea is to establish a
uniform bound on the size of any ``clock test'' in any truncated 
evaluation of $\cl t{\rho_\ell}$.  Once we do that, we can consider a
truncated evaluation with recursion depth greater than this bound.
In such a evaluation, either the recursion terminates normally
or the clock test will fail \emph{before}
any truncation axiom can be evaluated.  Either way, there are no
truncation axioms, so in fact we have an ordinary evaluation with
the given bound on its recursion depth.  Thus we will be able to
apply the Unfolding Lemma.

First we make an observation about evaluating $\crec$ terms.
The case of interest is a closure of the
form 
$\cl*{\crec (\bz^\ell)(\afflambda f.\lambda\vec v.t)t_1\dots t_k}\rho$ 
of base type.
The first (lowest) evaluation of~$t$
evaluates the closure
$\cl{t}{\extend{\rho_{\ell+1}}{v_i}{\cl{z_{\ell,i}}{\theta_{\ell,i}}}}$ 
where
$\cl{t_i}{\rho}\evalto\cl{z_{\ell,i}}{\theta_{\ell,i}}$.  Furthermore, if
$m\geq\ell$ then the evaluation of
$\cl{t}{\extend{\rho_{t,m+1}}{v_i}{\cl{z_{m,i}}{\theta_{m,i}}}}$
has the form
\begin{prooftree}
\AXC{}
\noLine
\UIC{$\cdots$}
\AXC{``$m+1 < \lh{z_{m+1,1}}$''}
\AXC{}
\noLine
\UIC{$\vdots$}
\noLine
\UIC{$\cl{t}{\extend{\rho_{t,m+2}}{v_i}{\cl{z_{m+1,i}}{\theta_{m+1,i}}}}\trevalto\cl z\theta$}
\BIC{$\cl{T_{m+1}}{\extend{\rho_{t,m+2}}{v_i}{\cl{z_{m+1,i}}{\theta_{m+1,i}}}}\trevalto \cl z\theta$}
\BIC{$\cl*{fs_1\dots s_k}{\extend{\extend{\rho_{t,m+1}}{v_i}{\cl{z_{m,i}}{\theta_{m,i}}}}{\vec y}{\vec{\cl {z'}{\theta'}}}}\trevalto \cl z\theta$}
\noLine
\UIC{$\ddots$}
\noLine
\UIC{$\cl t{\extend{\rho_{t,m+1}}{v_i}{\cl{z_{m,i}}{\theta_{m,i}}}}\trevalto \cl z\theta$}
\end{prooftree}
where:
\begin{itemize}
\item The $y$'s are the \keyw{let}-bound variables in~$t$;
\item $f\vec s$ is one of the complete applications of~$f$ in~$t$;
\item
$\cl{s_i}{\extend{\extend{\rho_{m+1}}{v_i}{\cl{z_{m,i}}{\theta_{m,i}}}}{\vec y}{\vec{\cl{z'}{\theta'}}}}\evalto \cl{z_{m+1,i}}{\theta_{m+1,i}}$
($f\notin\fv(s_i)$, so the evaluation of~$s_i$ cannot involve a
truncation axiom);
\item We assume that the evaluation of~$f\vec s$ hidden by the $\cdots$
does not use a truncation axiom to evaluate the $\crec$ term to
which $f$ evaluates.
\end{itemize}
This description of the evaluation is easy to prove by induction
on the shape of~$t$.
What we must do to prove the Termination Lemma is to get a handle
on the sizes of the values $z_{m,i}$ for $m\geq\ell$.

\begin{lem}
\label{parameter-bounds}
Suppose that $\typing{\Gamma_{\vec v}}{f\oftype\gamma}{t}{\base b}$
satisfies the ISA, $\rho\in\Env{\Gamma_{\vec v}}$,
$\varrho\in\Env{\tcden{\Gamma_{\vec v}}}$,
$\extend\rho{v_i}{\cl {z_{\ell,i}}{\theta_{\ell,i}}}\bddby\varrho$.
Consider any truncated evaluation of
$\cl{t}{\extend{\rho_{t,\ell+1}}{v_i}{\cl{z_{\ell,i}}{\theta_{\ell,i}}}}$.
Referring to the notation just introduced,
for any $m\geq\ell$, $\lh{z_{m,i}}\leq v_{ip}\xi_t^{m-\ell}\varrho$.
\end{lem}
\begin{proof}[\proofsketch]
The proof is by induction on $m-\ell$ with the base case given by
assumption.  For the induction step, we first bound~$\lh{z_{\ell+1,i}}$.
Here we need another claim about subterms of~$t$ as in the
proof of the Unfolding Lemma:
\begin{quotation}
Suppose that $\typing{\Gamma_{\vec v}^*}{f\oftype\gamma}{t^*}{\base b^*}$
is a subterm of~$t$ and take
$X^* = (P(\pot(f\star\vec{p^*}))+\cost(f\star\vec{p^*}),\,p^*(\pot(f\star\vec{p^*})))$
by the Decomposition Lemma so that $t^*\bddby X^*$.  Suppose
$\rho^*\in\Env{\Gamma_{\vec v}^*}$ is an extension of~$\rho$,
$\varrho^*\in\Env{\tcden{\Gamma_{\vec v}^*}}$ is an extension
of~$\varrho$, and 
$\extend{\rho^*}{v_i}{\cl{z_{\ell,i}{\theta_{\ell,i}}}}\bddby\varrho^*$.  
Then using notation analogous
to that just introduced, in the evaluation of
$\cl{t^*}{\rho^*_{t,\ell+1}}$, $\lh{z_{\ell+1,i}}\leq p_{ip}^*\varrho^*$.
\end{quotation}
The proof of the claim is by induction on the shape of~$t^*$ and
is by now routine.  Applying the claim to~$t$ we conclude that
$\lh{z_{\ell+1,i}}\leq p_{ip}\varrho$ and
so
$\extend{\rho}{v_i}{\cl{z_{\ell+1,i}}{\theta_{\ell+1,i}}}\bddby
\extend\varrho{v_i}{\val(p_{ip}\varrho)}$.
So for $m\geq\ell+1$ the induction hypothesis tells us that
$\lh{z_{m,i}}\leq v_{ip}\xi_t^{m-(\ell+1)}\extend\varrho{v_i}{\val(p_{ip}\varrho)}=
v_{ip}\xi_t^{m-\ell}\varrho$.
\end{proof}

\begin{thm}[Termination Lemma]
\label{termination-lemma}
Under the assumptions of Lemma~\ref{parameter-bounds},
% $\lh{z_{m,1}}\leq p_{1p}\varrho$ for all $m\geq\ell+1$.
% In particular,
$\rdp(\cl{t}{\extend{\rho_{\ell+1}}{v_i}{\cl{z_{\ell,i}}{\theta_{\ell,i}}}})\leq
(2+p_{1p})\varrho$.
\end{thm}
\begin{proof}
A key component of the One-step and $n$-step Lemmas of~\ATS\ (Lemmas~$44$
and~$45$) is that we can take~$p_{1p}$ such that $p_{1p}\xi_t = p_{1p}$
(this makes critical use of the restriction that if
$\base b_i\subtype\base b_1$ then $\base b_i$ is oracular).
Hence $v_{1p}\xi_t^d\varrho = p_{1p}\varrho$ for any $d\geq 2$.

Suppose we choose $d\geq 2$ such that $\ell+d-1\geq p_{1p}\varrho$.
Consider any truncated evaluation
of $\cl{t}{\extend{\rho_{t,\ell+1}}{v_i}{\cl{z_{\ell,i}}{\theta_{\ell,i}}}}$
of recursion depth~$d$.
Such an evaluation recursively evaluates
$\cl{t}{\extend{\rho_{t,m+1}}{v_i}{\theta_{m,i}}}$ for
$m=\ell,\dots,\ell+d-1$.  By Lemma~\ref{parameter-bounds}
we have that $\lh{z_{\ell+d-1,1}}\leq v_{1p}\xi_t^{d-1}\varrho =
p_{1p}\xi_t^{d-2}\varrho = p_{1p}\varrho \leq \ell+d-1$.
Thus either the evaluation terminates normally (i.e., the
evaluation of $\cl{t}{\extend{\rho_{t,m+1}}{v_i}{\cl{z_{m,i}}{\theta_{m,i}}}}$
does not recursively evaluate~$f$ at all for
some $\ell\leq m<\ell+d-1$) or one of the clock
tests fails, thereby terminating the evaluation.  Either
way we have a standard evaluation
of $\cl{t}{\extend{\rho_{t,\ell+1}}{v_i}{\cl{z_{\ell,i}}{\theta_{\ell,i}}}}$
of recursion depth~$\leq d$.  Taking
$d = (2+p_{1p})\varrho$ yields the theorem.
\end{proof}

\subsubsection{The Soundness Theorem.}

\begin{thm}
\label{soundness-prelim}
For every $\ATR$ term~$\GDtyping{t}{\tau}$ there is a 
$\tail(\tcden{\tau})$-safe t.c.\ denotation
$X$ of type~$\tcden\tau$ w.r.t.\ $\tcden{\Gamma;\Delta}$ such that $t\apprby X$.
\end{thm}
\begin{proof}
The proof is by induction on terms; for non-$\crec$ terms 
use Lemma~\ref{atr-minus-soundness-istep}.  
Let $s$ be the term
$\Gtyping{\crec (\bz^\ell)(\afflambda f.\lambda\vec v.t)}{\vec{\base b}\arrow\base b}$.
Suppose $\rho\in\Env\Gamma$,
$\varrho\in\Env{\tcden{\Gamma}}$, $\rho\apprby\varrho$.  
Since $\cl s{\rho}\evalto
\cl*{\lambda\vec v.T_\ell}{\rho_{t,\ell+1}}$ in one step, if
$\cl*{\lambda\vec v.T_\ell}{\rho_{t,\ell+1}}\bddby\chi$ then
$\cl s\rho\bddby\dally(1,\chi)$, so we focus on characterizing
such time-complexities~$\chi$.  Unwinding the definition of~$\bddby$,
we have have $\cl*{\lambda\vec v.T_\ell}{\rho_{t,\ell+1}}\bddby\chi$
if whenever $\cl{z_i}{\theta_i}\bddbypot p_i$ ($p_i$ is an arbitrary
potential here, not necessarily a polynomial), we have that:
\begin{enumerate}
\item $1\leq\cost(\chi),\cost(\pot(\chi)p_1),\dots,\tccost(\pot(\dotsc\pot(\pot(\chi)p_1)p_2\dotsc)p_{k-1})$.
\item $\cl{T_\ell}{\extend{\rho_{t,\ell+1}}{v_i}{\cl{z_i}{\theta_i}}}\bddby
\pot(\dotsc\pot(\pot(\chi)p_1)p_2\dotsc)p_k$.
\end{enumerate}
Since $\cl{z_i}{\theta_i}\bddbypot p_i$ we have that
$\extend\rho{v_i}{\cl{z_i}{\theta_i}}\bddby
\extend\varrho{v_i}{\val(p_i)}$.  
Let $\rho'$ and $\varrho'$ denote these extended environments.
By the Termination Lemma (Theorem~\ref{termination-lemma}) we have that
$\rdp(\cl t{\rho'_{t,\ell+1}})\leq
(2+p_{1p})\varrho$, where $p_{1p}$ is the $\potden{\base b_1}$-safe
polynomial given by the Decomposition Lemma (Theorem~\ref{decomposition-lemma})
for~$t$.  
By the Polynomial Unfolding Lemma (Corollary~\ref{solution-lemma-as-poly})
there is a $\base b$-safe polynomial $\phi(\vec v,d^{\potden{\base b_1}})$
such that
\[
  \cl{t}{\rho'_{t,\ell+1}}\bddby\phi(\vec v,p_{1p}+2)\varrho' =
  \pot(\dotsc\pot(\pot((\llambda_\star\vec v.\phi(\vec v,p_{1p}+2))\varrho)p_1)p_2\dotsc)p_k
\]
and hence
\[
  \cl{T_\ell}{\rho'_{t,\ell+1}}\bddby
  \pot(\dotsc\pot(\pot((\llambda_\star\vec v.\dally(8+v_{1p},\phi(\vec v,p_{1p}+2)))\varrho)p_1)p_2\dotsc)p_k.
\]
Since $\tccost(\llambda_\star x.X) = 1$ for any~$x$ and~$X$ and the
$\cl{z_i}{\theta_i}$ and $p_i$ were chosen arbitrarily, we conclude that
\[
  \cl*{\lambda\vec v.T_\ell}{\rho_{t,\ell+1}}\bddby
  (\llambda_\star\vec v.\dally(8+v_{1p},\;\phi(\vec v,p_{1p}+2)))\varrho.
\]
Since $\rho$ and $\varrho$ were chosen arbitrarily, we can therefore
conclude that
\[
  \crec(\bz^\ell)(\afflambda f.\lambda\vec v.t)\bddby
  \dally(1,\llambda_\star\vec v.\dally(8+v_{1p},\;\phi(\vec v,p_{1p}+2))),
\]
and by Propositions~\ref{safety-preservation} and~\ref{dally-preserves-safety},
this is a safe t.c.\ polynomial.
\end{proof}

\begin{defn}
For an $\ATR$ term $\GDtyping t\tau$ we define the \emph{time-complexity
interpretation of $t$}, $\tcden t$, to be the t.c.\ denotation
of Theorem~\ref{soundness-prelim}.
\end{defn}

\begin{cor}[Soundness for $\ATR$]
\label{soundness}
For every $\ATR$ term $\GDtyping t\tau$, $\tcden t$ is $\tail(\tcden\tau)$-safe
w.r.t.\ $\tcden{\Gamma;\Delta}$ and $t\apprby\tcden t$.
\end{cor}

\section{Second-order polynomial bounds}
\label{sec:polynomial-bounds}

Our last goal is to connect time-complexity polynomials to the usual
second-order polynomials of \citet{cook-kapron:sicomp96} and
show that any $\ATR$ program is computable in type-$2$ polynomial time.
The polynomial here will be in the lengths of the program's arguments,
and hence we need a semantics of \emph{lengths},
which lives inside the simple type structure over the time-complexity
base types.  We give a brief outline here, referring the reader
to Section~$2$ of~\ATS\ for full details.

For each $\ATR$-type $\sigma$ we define~$\lh\sigma$ by
\[
  \lh{\Nat_L} = \Tally_L
  \qquad
  \lh{\sigma\arrow\tau} = \lh\sigma\arrow\lh\tau.
\]
We are concerned primarily with two kinds of objects in these length-types:
the lengths of the meanings of $\ATR$ programs and
the meanings of second-order polynomials.  For the former,
recall that the interpretation of the~$\ATR$ base types is 
$K = \set{\bz,\bone}^*$;
for any $a\in K$, $\lh{a}$ is defined as expected and the length
of a function is defined as follows:

\begin{defn}
\label{defn:function-length}
If $f$ is a type-$1$ $k$-ary function, set
\[
  \lh{f} = \llambda n_1\dots n_k.\max\setst{\lh{f(v_1,\dots,v_k)}}{\forall i(\lh{v_i}\leq n_i)}.
\]
\end{defn}
The notion of length for objects of type-level~$\geq 2$
is much more difficult to pin
down; as we do not need it here, we omit any discussion of it.

With the notion of length in hand, we can give the definition
of $\tcden\alpha$ promised in Theorem~\ref{clm:atr-minus-soundness-prelim}:

\begin{defn}
If $\alpha^{(\base b_1,\dots,\base b_k)\arrow\base b}$ is an oracle
symbol, then
\[
  \tcden{\alpha^{(\base b_1,\dots,\base b_k)\arrow\base b}} =
  (1,\;\llambda n_1^{\potden{\base b_1}}(1,\;\llambda n_2^{\potden{\base b_2}}(\dots(1,\;\llambda n_k^{\potden{\base b_k}}(1\bmax\lh\alpha(\vec n),\lh\alpha(\vec n)))\dots))).
\]
\end{defn}

The \emph{second-order length polynomials} are defined by the typing rules in
Figure~\ref{fig:length-polys}; there is nothing surprising here, and the
intended interpretation is just as expected.  As with the time-complexity
types, we define
$\lh{\sigma}\shiftsto\lh{\tau}$ iff $\sigma\shiftsto\tau$ and
$\lh\sigma\subtype\lh\tau$ iff $\sigma\subtype\tau$.
\begin{figure}[t]
\begin{gather*}
\AXC{}
\UIC{$\Stctyping \eps {\Tally_\eps}$}
\DisplayProof
\quad
\AXC{}
\UIC{$\Stctyping {\mathbf{0}^n} {\Tally_\dmnd}$}
\DisplayProof
\\
\AXC{}
\UIC{$\tctyping {\Sigma,x\oftype\gamma} {x} \gamma$}
\DisplayProof
\\
\AXC{$\Stctyping p \gamma$}
\RightLabel{($\gamma\shiftsto\gamma'$)}
\UIC{$\Stctyping p {\gamma'}$}
\DisplayProof
\quad
\AXC{$\Stctyping p \gamma$}
\RightLabel{($\gamma\subtype\gamma'$)}
\UIC{$\Stctyping p {\gamma'}$}
\DisplayProof
\\
\AXC{$\Stctyping p {\base b}$}
\AXC{$\Stctyping q {\base b}$}
\BIC{$\Stctyping {p\bullet q}{\base b}$}
\DisplayProof
\quad
\AXC{$\Stctyping p {\base b}$}
\AXC{$\Stctyping q {\base b}$}
\BIC{$\Stctyping {p\bmax q}\base b$}
\DisplayProof
\\
\AXC{$\tctyping{\Sigma, x\oftype\lh\sigma}{p}{\lh\tau}$}
\UIC{$\tctyping\Sigma {\lambda x.p} {\lh{\sigma\arrow\tau}}$}
\DisplayProof
\quad
\AXC{$\Stctyping p {\lh{\sigma\arrow\tau}}$}
\AXC{$\Stctyping q {\lh\sigma}$}
\BIC{$\Stctyping {pq} {\lh\tau}$}
\DisplayProof
\end{gather*}
\caption{Typing rules for length polynomials.
The type~$\base b$ is a length base type,
$\gamma$ and $\gamma'$ are any length types,
and $\sigma$ and $\tau$ are any $\ATR$-types.
The operation~$\bullet$ is $+$ or $*$ and in this rule
$\base b$ is either $\Tally$ or $\Tally_{\dmnd_k}$ for some~$k$.
\label{fig:length-polys}}
\end{figure}
In these rules, a type-context~$\Sigma$ is an assignment of length-types
to variables.
For an $\ATR$ type-context $\Gamma;\Delta$ set
$\lh{\Gamma;\Delta} = \union_{(x\oftype\sigma)\in\Gamma;\Delta}\set{\lh{x}\oftype\lh\sigma}$
where for each $\ATR$ variable~$x$, $\lh{x}$ is a new variable symbol.

Our real concern is with closed $\ATR$ programs of the form
$\lambda\vec x.t$ where $t$ is of base type.  We know that
$\lambda\vec x.t\apprby \llambda_\star\vec x.\tcden{t} =
(1,\,\llambda_\star x_{1p}(\dots(1,\,\llambda_\star x_{kp}(P, p))\dots))$
where $P$ and $p$ are base-type polynomials over the potential
variables~$\vec x$.  Since the time-complexity polynomial calculus
is just a simple applied $\lambda$-calculus, it is strongly normalizing,
and so we can assume that the polynomials are in normal form.  Thus
we start with an analysis of time-complexity polynomials in normal form:

\begin{lem}
\label{tc-poly-normal-forms}
Suppose $\tctyping{x_1\oftype\potden{\sigma_1},\dots,x_k\oftype\potden{\sigma_k}}p\gamma$ 
is a t.c.\ polynomial in normal form.  Then $p$ has one of the following forms:
\begin{enumerate}
\item $\bz^n$ for some $n\geq 0$;
\item \label{item:potential}
$\pot(\pot(\dots\pot(\pot(vq_1)q_2)\dots)q_\ell)$ where $v$ is
either an oracle symbol or one of the $x_j$'s and each $q_i$
is in normal form and of potential type (this term is of potential type);
\item \label{item:cost}
$\tccost(\pot(\dots\pot(\pot(vq_1)q_2)\dots)q_\ell)$ where $v$ is
either an oracle symbol or one of the $x_j$'s and each $q_i$
is in normal form and of potential type (this term is of cost type);
\item $q_1*q_2$, $q_1+q_2$, $q_1\bmax q_2$ where each $q_i$ is in normal
form and of base type (this term is of base type);
\item $\alpha^{\tcden\sigma}$;
\item $(q_0,q_1)$ where $q_0$ is in normal form and of cost type and $q_1$
is in normal form and of potential type (this term is of time-complexity type);
\item \label{item:tc}
$\pot(\dots\pot(\pot(vq_1)q_2)\dots)q_\ell$ where $v$ is
either an oracle symbol or one of the $x_j$'s and each $q_i$
is in normal form and of potential type (this term is of time-complexity type).
\end{enumerate}
Note that (\ref{item:potential}) includes the special case $x_j$ when
$\sigma_j$ is a base type and in (\ref{item:potential}), (\ref{item:cost}),
and (\ref{item:tc}), $\ell$ may be strictly less than the arity of~$v$.
\end{lem}
\begin{proof}
By induction on the typing derivation.
\end{proof}

\begin{prop}
\label{lh-poly-from-potentials}
Suppose $p(x_1,\dots,x_k)$ is as in Lemma~\ref{tc-poly-normal-forms},
$\alpha_i^{\sigma_i}$ an oracle symbol for $i=1,\dots,k$.  Then
$p(\pot(\tcden{\alpha_1}),\dots,\pot(\tcden{\alpha_k}))$ is
a second-order polynomial in $\lh{\alpha_1},\dots,\lh{\alpha_k}$.
\end{prop}

Combining the Soundness Theorem (Corollary~\ref{soundness}) 
with Proposition~\ref{lh-poly-from-potentials} yields:

\begin{thm}
\label{poly-time-computability}
If $\typing{\underline{~}}{\underline{~}}{t}{\tau}$, then
$t$ is computable in type-2 polynomial time.
\end{thm}

A word of caution in interpreting this result is in order.
The \emph{basic feasible functionals} of
\citet{mehlhorn:stoc74} and
\citet{cook-urquhart:fca} are an extension of polynomial-time functions
to higher type.  They live in the full (set-theoretic) type structure and for
type-level~$\leq 2$ are defined as follows.  The basic model is an
oracle Turing machine with function oracles, and the cost of an
oracle query is the length of the answer.  A 
functional~$F(f,x)$ of type-level~$\leq 2$
is \emph{basic feasible} if it is computed by such an oracle Turing machine
with oracle~$f$ in time $p(\lh f,\lh x)$, where $p$ is a second-order
polynomial (this is the characterization of
\citet{cook-kapron:sicomp96}; 
\citet{ignjatovic-sharma:tocl02} give a similar characterization for
unit-cost oracle queries).  
Now, $\ATR$ is \emph{not} interpreted in the
full type structure but rather in the
well-tempered semantics discussed
in Section~\ref{sec:time-complexity-semantics}.  Thus,
we have not quite yet characterized the
basic feasible functionals.  
However, on $\ATR$-types that are both strict and predicative
(see Definition~\ref{defn:predicative-etc}), the well-tempered
semantics agrees with the full type structure (recalling the
discussion after Definition~\ref{defn:predicative-etc}, the relevant
point here is that no restrictions are made on function spaces of
strict and predicative type).  Thus we conclude:

\begin{thm}
If $\typing\emptyctx\emptyctx t\tau$, all variables of~$t$ are of
strict and predicative type, and $t$ contains no oracle symbols,
then $t$ defines a basic feasible functional.
\end{thm}

In fact, some $\ATR$ programs compute function(al)s that
are not basic feasible but are nonetheless second-order polynomial-time
computable according to Theorem~\ref{poly-time-computability}.  
For example, consider the following
$\ATR$ program for the primitive recursion on notation combinator
(roughly, \keyw{foldr} for binary strings):
\begin{lstlisting}
val prn : $((\Nat_\eps\arrow\vec{\base b}\arrow\Nat_{\Box_1}\arrow\Nat_{\Box_1})^2,\Nat_\eps)\arrow(\Nat_\eps\arrow\vec{\base b}\arrow\Nat_{\Box_1})$
    fn f0, f1, a $\Rightarrow$
        fn x $\vec y$ $\Rightarrow$ letrec F : $\Nat_{\Box_1}\arrow\Nat_\eps\arrow\Nat_{\Box_1}$ =
            fn b x' $\Rightarrow$ if x' then if $\test_0$x' then f0 (d x') $\vec y$ (F b (d x'))
                                  else f1 (d x') $\vec y$ (F b (d x'))
                        else a
        in F $\cons_0$x x end
\end{lstlisting}
This combinator is not basic feasible, because in the full type structure
it could be applied to arguments with non-trivial growth rates, and this
would lead out of the realm of feasibility.  
However, in $\ATR$ the types of the arguments control the growth rates of
the functions to which it is applied (specifically, the type of the function
argument ensures that it has a
``small'' growth rate in terms of the size of the recursive call).  
Thus we can have our cake and
eat it too:  we can define natural programming combinators 
(like $\mathit{prn}$),
but the type system will keep us from using them in a
way that results in infeasible computations.

\section{Concluding remarks}
\label{sec:concl}
\suppressfloats

In~\ATS\ we introduced the formalism~$\ATR$ which captures 
the basic feasible functionals at type-levels~$\leq 2$.  In the current
paper we have
extended the formalism to include a broad range of affine recursion schemes 
(plain affine recursive definitions)
that allow for more
natural programming and demonstrated the new formalism by implementing
lists of binary strings and insertion- and selection-sort. 
We have extended the original time-complexity semantics
of \ATS\ to handle the more involved programs expressible via
plain affine recursion and shown that these new programs do
not take use out of the realm of feasibility.
We conclude by indicating some possible extensions and future 
research directions:

\paragraph{Branching recursion.}
This paper has focused on affine (one-use) recursions, and of course there
are feasible algorithms that do not fit this mold.
Especially germane to the examples of this paper are sorting algorithms
such as merge-sort and quick-sort that are based on branching recursions.
Let us consider the latter to see some of what would be involved in 
adding branching recursion to an $\ATR$-like language.
Here is a functional version of quick-sort over lists:
\begin{lstlisting}
val quicksort =
    fn xs $\Rightarrow$ letrec qsort =
        fn ys $\Rightarrow$ if (length ys) $\leq$ 1 then ys
                 else let val (pivot,small,big) = partition ys
                      in append (qsort small) (cons pivot (qsort big)) end
     in qsort xs end
\end{lstlisting}
We assume that \lstinline!small! is the list of items in
\lstinline!ys! with values $\leq \lstinline!pivot!$
(excluding the pivot item itself), and \lstinline!big! is the list of items in
\lstinline!ys! with values $> \lstinline!pivot!$.

The tightest upper bounds on the sizes of the individual
arguments are
$|\lstinline!small!| < |\lstinline!ys!|$ and 
$|\lstinline!big!|<|\lstinline!ys!|$, and this only allows us to
extract exponential upper bounds on the run-time of this definition.
In order to establish a polynomial run-time bound
one also needs to know that that the arguments of the two
branches of the recursion satisfy the \emph{joint} size restriction
$|\lstinline!small!| + |\lstinline!big!| < |\lstinline!ys!|$.
It is hard to see how to gracefully assert this sort of joint size
bound using $\ATR$-style types and combinators.  Another problem is
that in a recursive definition, it may be difficult to know which of
the various recursive calls can together form a set of branching
calls, and hence it may be difficult to know what sets of joint
size constraints one needs to satisfy to guarantee a polynomial
run-time.

Rather than attempting to handle general feasible branching
recursions, we propose investigating combinators that express
particular flavors of branching recursions that work well with
$\ATR$-style types and deal with the problems noted above.  Here is a
reworked version of quick-sort using a possible such combinator,
inspired by Blelloch and colleagues' work on 
the parallel programming language
NESL~\citep{blelloch-et-al:jpdc94,blelloch:cacm96}:
\begin{lstlisting}
val quicksort =
    fn xs $\Rightarrow$ letrec qsort =
        fn ys $\Rightarrow$ if (length ys) $\leq$ 1 then ys
                 else let val (zs, part_idx) = partition ys
                      in concat (map$'$ qsort zs [part_idx, part_idx+1]) end
    in qsort xs end
\end{lstlisting}
Here we assume that \lstinline!partition! is defined so that
\lstinline!zs! is a permutation of
\lstinline!ys! such that $\lstinline!zs[i]! \leq \lstinline!zs[part_idx]! <
\lstinline!zs[j]!$ for
any $i\leq\lstinline!part_idx!<j$,
and \lstinline!map$'$ f vs [i, j]! maps \lstinline!f! over 
\lstinline![vs[0..i-1], vs[i..j-1], vs[j..(length zs)-1]]!.
% and \lstinline!map$'$ f vs [i, j, k]! map$'$s \lstinline!f! over 
% \lstinline![vs[i..j-1], vs[j..k-1]]!.
Notice that in this definition, \lstinline!qsort! occurs affinely
(modulo \lstinline!map$'$!) and the aggregate data to the branching
recursion (i.e., \lstinline!zs! in the \lstinline!map$'$! expression) 
occurs in one
place where typing has a chance of constraining its size.  Based on
this, we claim it is quite plausible that a combinator like
\lstinline!map$'$! can be integrated into $\ATR$, and thanks to the work on
NESL we know that such a combinator can express a great many useful
divide-and-conquer recursions.  In fact, NESL uses a parallel
\lstinline!map$'$! combinator, so
using the NESL work one could do a straightforward
static analysis of $\ATR+\lstinline!map$'$!$-programs to extract
bounds on their \emph{parallel} time complexity.
This would fit in very nicely with recent work of
\citet{chakravarty-et-al:damp07} on data-parallel Haskell.

\paragraph{Lazy $\ATR$.}
A version of~$\ATR$ with lazy evaluation
    would be very interesting, regardless of whether the constructors are
	strict or lazy (yielding streams).  There are many technical challenges in
    analyzing such a system but we expect that the general outline
    will be the approach we have used in this paper.  
    Of course one can implement streams in
    the current call-by-value setting in standard ways (raising the
    type-level), 
    but a direct lazy implementation of streams is likely to be
    more informative.
    We expect the analysis of such a lazy-$\ATR$
    to require an extensive reworking
    of the various semantic models we have discussed here and in~\ATS.

\paragraph{Real-number algorithms.}
$\ATR$ is a type-$2$ language, but here we have focused on type-$1$
    algorithms.  We are interested in type-$2$ algorithms, specifically 
	in real-number algorithms
	as discussed in, e.g., \citet{Ko:Comp-Compl-Real-Fns}, where real numbers
	are represented by type-$1$ oracles.
    This can be done in either a call-by-value setting in which 
	algorithms take a
    string of length~$n$ as input and return something like an $n$-bit
    approximation of the result,
	or a lazy setting in which the algorithm returns
    bits of the result on demand.
	Combined with lazy constructors, the latter would allow us to
	view real numbers themselves as streams; in particular, since
	real numbers would be base-type objects, we could look at
	operators on real functions.

\appendix

\section{Equivalence of the operational semantics and the abstract machine 
semantics of~\ATS}
\label{app:equiv-semantics}

Here we sketch the proof of equivalence between the abstract-machine
semantics for~$\ATR$ in~\ATS\ and the evaluation-derivation semantics
we have used here.
We refer the reader to \ATS\ for a detailed
definition of the abstract machine.  The abstract machine semantics
works with \emph{configurations} of the form $\cfg* t \rho \kappa$, where
$t$ is an expression, $\rho$ an environment, and $\kappa$ a
(defunctionalized) continuation, and defines a \emph{transition relation}
$c\transto c'$ between configurations.  Continuations are defined as
a sequence of keywords, expressions, and environments, always ending
in the keyword~$\cont{halt}$.  If $\kappa$ and $\kappa'$ are two
continuations, we define $\kappa\kappa'$ to be the continuation obtained
by deleting the keyword $\cont{halt}$ from $\kappa$ and then concatenating
$\kappa'$ to the result.  For configurations $c$ and $c'$ we write
$c\transto^n c'$ if $c=c_0\transto c_1\transto\dots\transto c_{n}=c'$ and
$c\transto^* c'$ if $c\transto^n c'$ for some~$n$.  In the following,
$z$ denotes a value.

\begin{lem}
\label{clm:extend-cont}
If $\cfg* t\rho{\kappa_0}\transto^n\cfg* z\theta{\kappa_1}$, and $\kappa'$ is
any continuation, $\cfg* t\rho{\kappa_0\kappa'}\transto^n
\cfg* z\theta{\kappa_1\kappa'}$.  In particular,
if $\cfg t\rho{\cont{halt}}\transto^n\cfg z\theta{\cont{halt}}$, then
for any continuation~$\kappa$,
$\cfg* t\rho{\kappa}\transto^n\cfg* z\theta{\kappa}$.
\end{lem}

\begin{prop}
If $\cl t\rho\evalto_n\cl z\theta$ then
$\cfg t\rho{\cont{halt}}\transto^{m}\cfg z\theta{\cont{halt}}$ for
some $m\leq 3n$.
\end{prop}
\begin{proof}
By induction on the height of the derivation.  Lemma~\ref{clm:extend-cont}
allows us to make use of the induction hypothesis.
\end{proof}

\begin{lem}
\label{clm:trans-init-seg}
If $\cfg* t\rho{\kappa_0}\transto^*\cfg* z\theta{\kappa_1}$, then the
transition sequence has an initial segment of the form
$\cfg* t\rho{\kappa_0}\transto^n\cfg* {z'}{\theta'}{\kappa_0}$ for some value
$\cl {z'}{\theta'}$ such that $\cl t\rho\evalto_n \cl{z'}{\theta'}$.
\end{lem}
\begin{proof}
By induction on the length of the transition sequence.
\end{proof}

\begin{prop}
If $\cfg t\rho{\cont{halt}}\transto^n\cfg z\theta{\cont{halt}}$, then
$\cl t\rho\evalto_n\cl z\theta$.
\end{prop}
\begin{proof}
By Lemma~\ref{clm:trans-init-seg}, there are $m$ and $\ell$ such that
the given transition sequence has the form
$\cfg t\rho{\cont{halt}}\transto^m\cfg {z'}{\theta'}{\cont{halt}}\transto^\ell
\cfg z\theta{\cont{halt}}$.  Since $z'$ is a value, there are no transitions
that start from $\cfg {z'}{\theta'}{\cont{halt}}$, and so we conclude
that $\ell=0$ and hence that $\cl {z'}{\theta'}=\cl z\theta$.  And by
Lemma~\ref{clm:trans-init-seg}, $\cl t\rho\evalto_m\cl {z'}{\theta'}=\cl z\theta$.
\end{proof}

\bibliographystyle{abbrvnat}
\bibliography{master}

\end{document}